\documentclass[]{article}

\usepackage{babel}
\usepackage{pifont}
\usepackage[margin=3cm]{geometry} 
\usepackage[authoryear]{natbib}
\usepackage{threeparttable}

\usepackage{quattrocento}

\usepackage{setspace}

\usepackage{fancyhdr}
\usepackage[compact]{titlesec}

\usepackage{sectsty}
\allsectionsfont{\fontfamily{phv}\selectfont}

\usepackage{amsmath}
\usepackage{amssymb}
\usepackage{bm}
\usepackage{caption}
\usepackage{graphicx}
\usepackage{cancel}
\graphicspath{{./Figures/}}

\usepackage{adjustbox}
\usepackage{makecell}

\usepackage{enumitem}
\setlist{nolistsep}
\usepackage[symbol]{footmisc}

\usepackage{siunitx}

\renewenvironment{abstract}{%
	\hspace{0.025\linewidth}\begin{minipage}{0.95\textwidth}
		\rule{\textwidth}{1pt}\small\selectfont}
	{\vspace{-0.5em}\par\noindent\rule{\textwidth}{1pt}\end{minipage}\vspace{1em}}
\makeatletter
\renewcommand{\maketitle}{\bgroup\setlength{\parindent}{0pt}
	\thispagestyle{empty}
	\begin{flushleft}
		{\bf \fontfamily{phv}\selectfont \LARGE \@title}
		
		{\bf \fontfamily{phv}\selectfont \@author}
		
		\fontfamily{phv}\selectfont \small \@date
	\end{flushleft}\egroup
}
\makeatother
\title{Recovering head and flux distributions at the sediment-water interface for arbitrary, transient bedforms by inversion of photographic time series}

\author{Yoni Teitelbaum$^1$, Shai Arnon$^1$, Aaron Packman$^2$, Scott K. Hansen$^1$\footnote[2]{Corresponding author. Email: skh@bgu.ac.il}}
\date{$^1$Ben-Gurion University of the Negev, Israel\\$^2$Northwestern University, USA}
\begin{document}
	\doublespacing
	\maketitle
	\onehalfspacing

\begin{abstract}
Existing works that predict bedform-induced hyporheic exchange flux (HEF) typically either assume a simplified streambed shape and corresponding sinusoidal head distribution or rely on costly computational fluid dynamics simulations. Experimental data have been lacking for the formulation of \textit{a priori} prediction rules for hydraulic head and flux distributions induced by spatiotemporally heterogeneous natural bedforms because it has not previously been feasible to determine these in the laboratory. We address this problem, presenting a noninvasive technique for regularized inversion of photographic time series of dye front propagation in the hyporheic zone, compatible with arbitrarily-shaped, generally transient bedforms. We employ the technique to analyze three bench-scale flume experiments performed under different flow regimes, presenting a new data set of digitized bed profiles, corresponding head distributions, and dye fronts. To our knowledge, this is the first such data set collated for naturally-formed sand bedforms. 
\end{abstract}

\section{Introduction}

In the study of streambeds, the sediment-water interface (SWI) is of great importance to both geology and hydrology. In flowing systems with fine sediments (0.1-10 mm), the bed arranges itself in characteristic structures known as bedforms \citep{Ashley1990,Kennedy1969}. Interaction between surface water flow and bedforms result in a spatially varying pressure distribution along the sediment-water interface (SWI), which drives Darcy flow in the subsurface, beneath the SWI \citep{Harvey1996,Stonedahl2010,Winter1998}. The exchange between surface and subsurface water is referred to as hyporheic exchange flux (HEF), and the region of the sediment where HEF occurs is called the hyporheic zone (HZ) \citep{Boano2014}.

An early discussion of the flow of stream water in and out of gravel streambeds was given by \cite{Vaux1968}, who proposed that a bedform's geometry induced its pressure distribution and also that bedform head was linearly related to elevation, a concept that has been refined by more recent authors. Bedform-induced HEF was first observed experimentally in 1987 by \cite{Thibodeaux1987a} and simulated numerically by \cite{Savant1987}. By this time, the forces at play between the surface water and the SWI had already been the focus of study by the sediment transport community for several decades \citep{Bennett1995}.

Several key studies since \cite{Vaux1968} have directly considered the relationship between bedform shape and head or pressure distribution (Table \ref{table:table-prev-exps}). For example, \cite{Raudkivi1963} carried out flume experiments to formulate a physical explanation for the formation of ripples in a sandy streambed. In this course of this study, he used pressure taps to measure pressure along a sheet-metal replica of a naturally-formed bedform. However, no formula for the spatial distribution of pressure over the bedform was given. \cite{Vanoni1967} investigated the head gradient across naturally-formed sand bedforms in flume experiments. This was accomplished by spraying the naturally-formed bedforms with an adhesive spray and then installing pressure taps flush with the bedform surface. They formulated a logarithmic expression for form drag across a bedform as a function of bedform shape parameters. \cite{Vittal1977} carried out flume experiments over bedforms constructed out of plywood and coated with sand having $d_{50}$ = 0.6mm. Skin shear was measured using Preston tubes, and pressure was measured using an inclined manometer. This work disproved the prevailing assumption at the time that skin shear is constant along the face of the bedform. In the process, \cite{Vittal1977} also measured and formulated a relation for pressure exerted along the bedform by the surface water flow. \cite{vanMierlo1988} measured flow velocities and pressure over sand-coated concrete bedforms in a flume. Their measurements of pressure induced along the bedform are similar to those found by other works, with a local maximum approximately halfway along the stoss face of the bedform, and local minima in the vicinity of the bedform crest. \cite{Fehlman1985} was the first to experimentally confirm the conjecture that the total force exerted by the surface flow on 2D bedforms is the sum of the form drag (pressure differential across the bedform) and the skin shear along the bedform surface. This was accomplished by measuring all three quantities independently using a force balance, surface pressure taps, and a shear probe mounted around plastic bedforms. Additionally, this work generalized the relation for pressure exerted by the flow along the face of the bedform reported by \cite{Vittal1977}. In some runs, the bedforms were smooth, while in others they were coated with 1.75 mm (median grain diameter) sand to simulate streambed roughness.

While previous studies had largely focused on geomorphology as the driving force behind HEF, \cite{Sawyer2011} showed using experiments and modeling that in-stream woody debris also creates a head gradient which drives HEF. They conducted flume experiments in which they represented channel-spanning logs by mounting PVC pipes over a flat sand bed (median grain diameter 2.4 mm). Pressure sensors were installed near and below the SWI. pore water flow was visualized with dye tracer tests, and salt tracer tests were used to measure HEF. This work showed that debris generates HEF at levels comparable to bedforms at the reach scale. Similar results were attained by \cite{Huang2023} for porous logjams in flume experiments with hydrogel-bead sediment and porous logs represented by piled acrylic cylinders.

Prior to \cite{Savant1987}, HEF had been modeled by the Transient Storage model (TSM) \citep{Bencala1983}. The TSM is a 1D model of downstream solute transport which ascribes solute retardation to flux in and out of a stream-adjacent, black-box storage zone.  \cite{Elliott1997} adapted the observations of \cite{Fehlman1985} to a 2D analytical model describing pore water flow paths with a sinusoidal boundary condition like that of \cite{Toth1963}, while corroborating the model with experimental results of their own \citep{Elliott1997a}. They used sand bedforms and measured flux via tracer tests, but they did not measure head gradients along the SWI. \cite{Cardenas2007b} computed the pressure distribution over bedforms using the Reynolds-Averaged Navier Stokes (RANS) method. This pressure distribution was then used as the top boundary condition to solve the groundwater flow equation within the pore water. \cite{Janssen2012} corroborated the CFD-calculated pressure boundary condition against results from flume experiments that they conducted. They shaped a sediment bed consisting of 0.174 mm (mean grain diameter) sand into triangular bedforms. Head at the SWI was measured by pressure sensors installed flush with the sediment surface along the flume midline. Surface flow was measured by Particle Image Velocimetry (PIV). pore water flow was assessed by injecting and measuring fluorescent tracer in the sediment bed. More recently, \cite{Li2020a} implemented a numerical model featuring full coupling between surface and subsurface flow.

Other works have addressed the relationship between hydraulic head and more complex streambed formations. \cite{Marion2002} showed using salt tracer experiments over naturally-formed bedforms in a flume that the model developed by \cite{Elliott1997} is less accurate for bedforms that protrude significantly above the mean bed elevation and have a large height relative to flow depth. \cite{Jin2022} used simulations to show that for a given set of bedform dimensions, a wavelike bedform shape induces greater HEF than a triangular one. Several works used numerical simulations to show that in-stream cobbles and vegetation also drive HEF \citep{Lee2022,Yuan2021,Lv2022}. Marzadri et al. (\citeyear{Marzadri2010}) and Huang and Chui (\citeyear{Huang2022}) used experiments and modeling to show that alternate bars in streams also contribute to HEF, and simulations by \cite{Cardenas2009} showed that streambed sinuosity does as well. Cardenas (\citeyear{Cardenas2008}) and Lee et al. (\citeyear{Lee2020}) used numerical simulations to examine HEF under landscapes containing multiple scales of features, from superimposed ripples to the watershed. These works found that each scale of feature has its own significant effect on pore water flow and residence times. Additionally, Chen et al. (\citeyear{Chen2015}, \citeyear{Chen2018}) used simulations of flow over 3D bedforms to show that using 2D bed profiles alone does not capture the full extent of HEF in the streambed. In addition to the aforementioned CFD-based approaches, some hybrid spectral approaches have been proposed \citep{Worman2006, Worman2007, Stonedahl2010} which superpose Elliott and Brooks's sinusoidal solutions for irregular morphologies. To our knowledge, these spectral approaches have been corroborated \citep{Stonedahl2010} only in a global sense by matching uptake from salt tracer tests.

In general, the sinusoidal approach of \cite{Elliott1997} remains state-of-the art for \textit{a priori} prediction of bedform head distributions in the absence of CFD. Predicting head distributions for general, spatially-heterogeneous bedforms, such as those observed in nature, is an important open problem, for which relevant experimental data is needed.

In the experimental works above, moving and even naturally-formed bedforms are significantly under-represented (Table \ref{table:table-prev-exps}). All of the experimental studies of force exerted by the stream water on the SWI used static, regularly shaped bedforms, and all but Janssen et al. (\citeyear{Janssen2012}) used bedforms that were made of artifical materials such as metal, plaster, stabilized sand, or wood. Additionally, almost all of the spatial models of HEF listed above used a static bed with a simplifed domain shape. \cite{Teitelbaum2022} used a model whose domain is translated horizontally in connection with bedform motion, but the translated domain shape was regular and periodic, and its motion was constant. Stonedahl etl al. (\citeyear{Stonedahl2010}) and W{\"o}rman et al. (\citeyear{Worman2006}) used an irregular domain shape, but the results in that work were verified only at a global and not a local level. Cardenas (\citeyear{Cardenas2008}) and Lee et al. (\citeyear{Lee2020}) simulated irregular domain shapes with no experimental verification.

In this work we recover the hydraulic head distributions induced by surface water flow over and through naturally- and manually-formed sand bedforms in an ensemble of flume experiments performed using a range of flow velocities. To our knowledge, no dataset has previously been reported for dynamically evolving, generally non-stationary natural sand bedforms at the higher flow rates we considered. This recovery was enabled by non-intrusive optical velocimetry employing time lapse photographs of dye tracer propagation in a glass-sided flume, using gradient-descent inversion on a physically regularized objective function. The inverse computational method we used to attain these results is also a substantial contribution of this work, as it does not involve physical instrumentation of the bed, and thus can be used for temporally evolving real sand bedforms.

The rest of this work is divided as follows. In Section \ref{subsec:dyetracertests} we outline a set of dye tracer flume experiments. In section \ref{sec: head-flux} we outline the theory that allows for recovery of head and flux distributions from dye front dynamics in the experiments. In Section \ref{sec: implementation}, we discuss the details of computer implementation, including numerical discretization, validation, and practical considerations for dealing with imperfect data. In Section \ref{sec:results} we outline our results. We summarize our experimental data set and present recovered head distributions recovered for three separate experiments. Finally, in Sections \ref{sec:discussion} and \ref{sec:conclusion}, we discuss insights to be drawn from the results, limitations of the current approach, and promising directions for future research.

\begin{table}
	\centering
	\begin{threeparttable}
	\begin{tabular}{lcccc}
		\hline
		\multicolumn{1}{l}{\bfseries Source} & \multicolumn{1}{c}{\bfseries \makecell{Bedform \\ description}} & \multicolumn{1}{c}{\bfseries \makecell{Bedform \\ wavelength (m)}} & \multicolumn{1}{c}{\bfseries \makecell{Bedform \\ height (m)}} & \multicolumn{1}{c}{\bfseries \makecell{Flow \\ velocity (m/s)}} \\ \hline
		\hline
		\cite{Raudkivi1963} & \makecell{Metal sheet (smooth)} & 0.386 & 0.022 & 0.299 \\ \hline
		\makecell[l]{Vanoni and\\ Hwang [\citeyear{Vanoni1967}] } & \makecell{Sand (chemically stabilized, \\ naturally evolved)} & \makecell{ 0.156 \\ 0.162} & \makecell{0.016 \\ 0.014} & \makecell{0.228 \\ 0.381} \\ \hline
		\cite{Vittal1977} & \makecell{Plywood (smooth) \& \\ Plywood (sand coated)} & \makecell{0.15 \\ 0.3 \\ 0.45 \\ 0.6} & 0.03 & [not reported] \\ \hline
		\cite{Fehlman1985} & \makecell{Plastic (smooth) \& \\ Plastic (sand coated)} & 0.914 & 0.137 & 0.147-0.653 \\ \hline
		\makecell[l]{van Mierlo and\\ de Ruiter [\citeyear{vanMierlo1988}]} & \makecell{Concrete (sand coated)} & 1.6 & 0.08 & \makecell{0.394 \\ 0.513} \\ \hline
		\cite{Janssen2012} & \makecell{Sand (hand shaped)} & 0.2 & 0.02 & \makecell {0.07 \\ 0.12} \\ \hline
		This work & \makecell{Sand (hand shaped) \& \\ Sand (naturally evolved)} & \makecell{ 0.15 \\ 0.20 \\ 0.37}  & \makecell{0.02 \\ 0.024 \\ 0.018} & \makecell{0.123 \\ 0.22 \\ 0.31} \\ \hline
	\end{tabular}
	
	
	\caption{Experimental works that have measured pressure and/or head distributions induced on bedforms by stream flow. \label{table:table-prev-exps}}
	\end{threeparttable}
\end{table}

\section{Dye tracer tests}
\label{subsec:dyetracertests}

	Three dye tracer experiments were conducted in a glass-sided flume containing natural sand, at three different mean water velocities, named A, B, and C, in order of increasing velocity. In all the experiments, dye was introduced into the surface water which was recirculated throughout the flume. HEF progressively introduced the dye to the sand, creating visibly evolving lobes of dye in the hyporheic zone. The flume employed was 640 cm long by 29 cm wide, and a Nikon DSLR D5300 camera was mounted in a fixed position facing its side wall, viewing a portion of the bedforms and the subsurface. Water was filled over a packed sediment bed of homogeneous sand, and surface water was recirculated over the sediment bed using a centrifugal pump. The sand had mean grain diameter 384 µm, hydraulic conductivity K = 0.12 cm/s, and porosity 0.33. In Experiment A, the sediment was manually shaped into regular, repeated bedforms of length 15 cm and height 2 cm using a stencil, while in Experiments B and C the bedforms were formed naturally, starting from an initially flat sand pack. 25 g of Brilliant Blue dye dissolved in 5 L distilled water was added to the flume surface water prior to the experiments, which was quickly dispersed throughout the surface water. The camera collected photos of the dye plume in the subsurface at regular intervals, with the evolution. The key parameters describing the three experiments are summarized in Table \ref{table_exp_img_params}.  

	\begin{figure}
		\centering
		\includegraphics[width=0.9\linewidth]{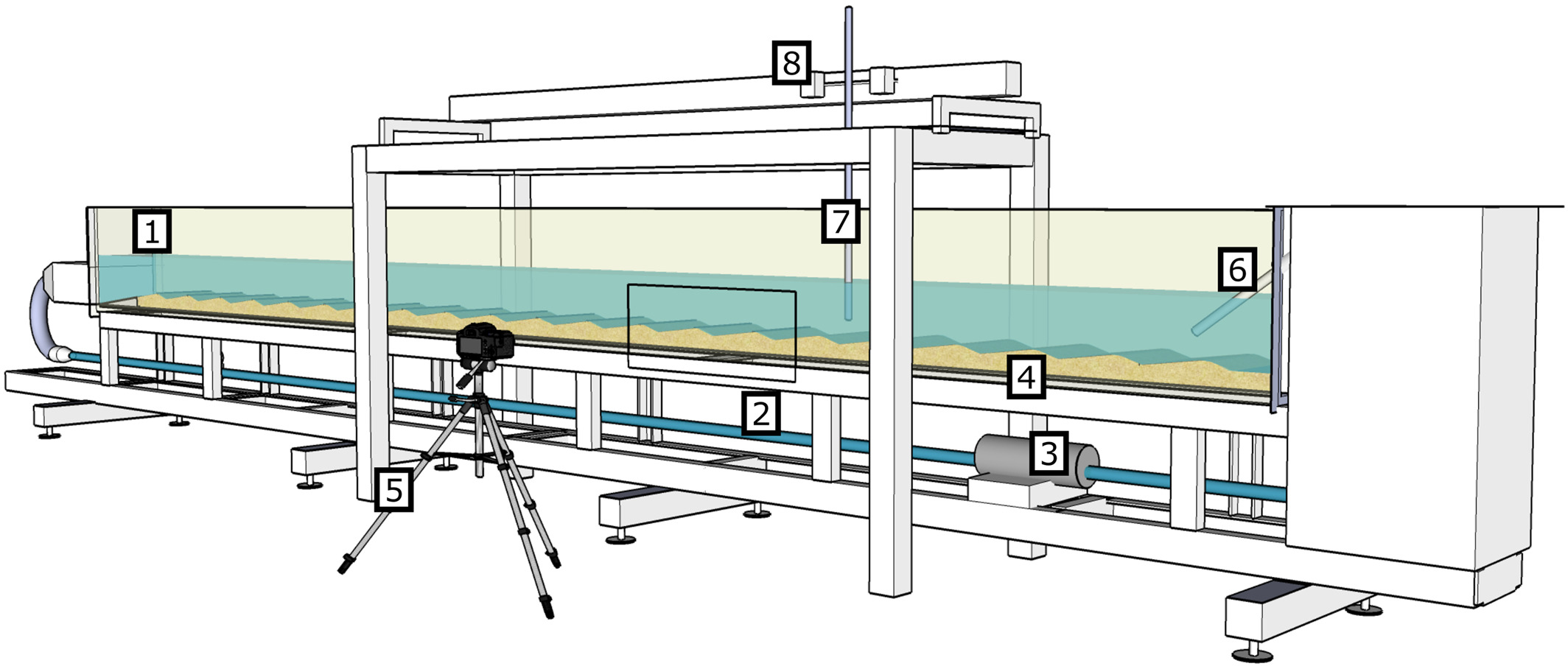}
		\caption{Rendering (original from \cite{Teitelbaum2021}) of large-scale flume apparatus used in dye tracer experiments. Relevant for this work are the recirculated dye-filled water [1] and the camera setup [5] which captures a small portion of the subsurface under controlled physical and lighting conditions.}
		\label{fig: setup}
	\end{figure}

	Experiment A was part of the analyses conducted by \cite{Fox2014}, where the full description of the experimental setup can be found. Experiment B has not previously been published in any form, and Experiment C was conducted as part of the experiments of \cite{Teitelbaum2021}, where more details about the experimental procedures can be found.
	
\begin{figure}
		\centering
		\includegraphics[width=1.0\linewidth]{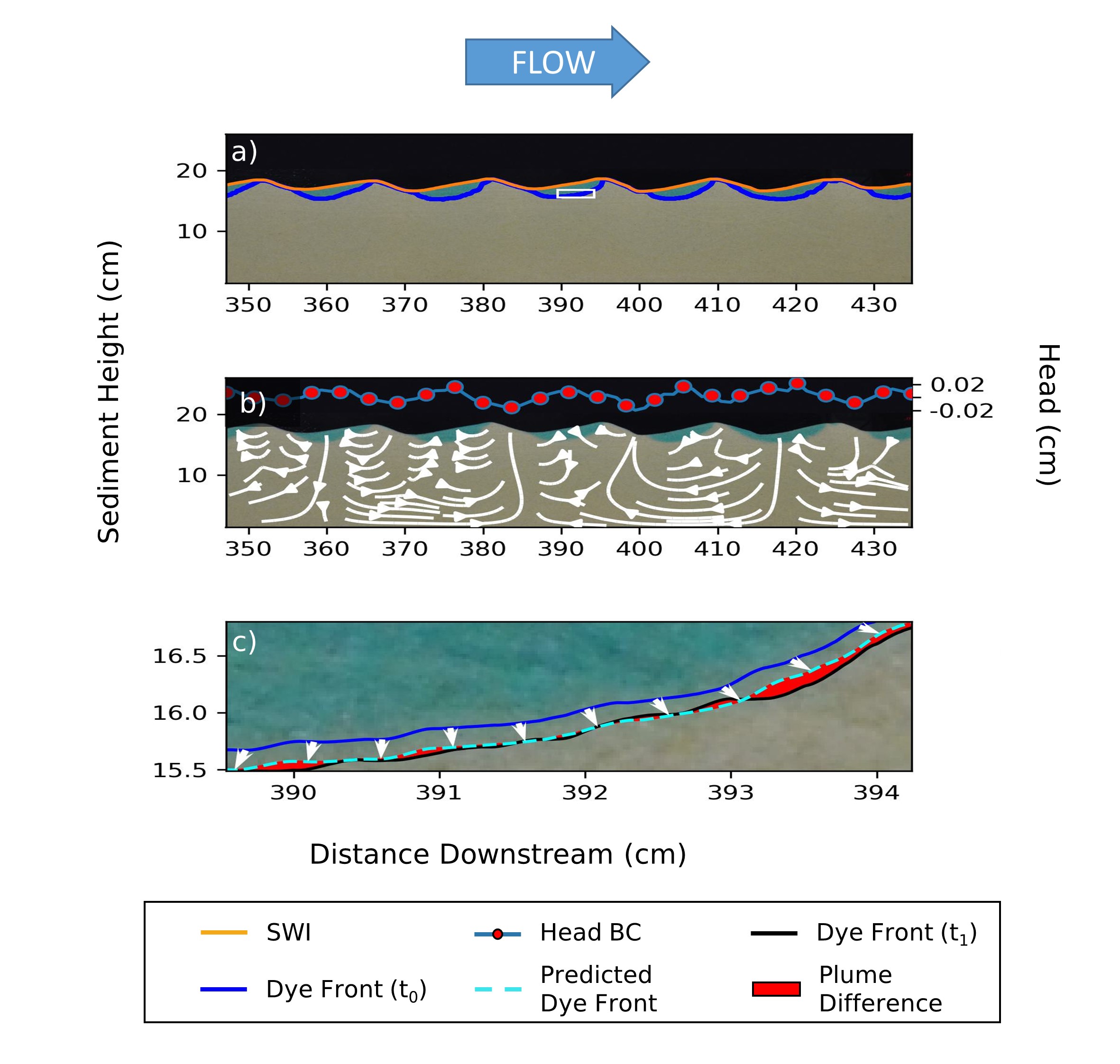}
		\caption{Illustration of the optimization procedure for a given time step. a) The SWI and dye front extracted from a time lapse photo using image processing. b) A candidate head BC with streamlines (white) showing the corresponding flow field. Red circles represent interpolation points. c) The observed dye fronts $\partial \Omega_t$ and $\partial \Omega_{t+n}$, the predicted dye front $\partial \hat{\Omega}_{t+n}$, and the plume difference. White rectangle in a) delineates the region shown in panel c).}
		\label{fig:optimizationpertimestepwithimages}
	\end{figure}

	The computer vision software libraries OpenCV \citep{Bradski2000} and scikit-image \citep{scikit-image} were used to process the images in Python. Two pieces of information were extracted for each image: the shape of the SWI, and the location of the dye front (Figure \ref{fig:optimizationpertimestepwithimages}). Segments of the dye front near the domain side and top boundaries were omitted due to data quality issues.
	
	\begin{table}
		\centering
		\begin{tabular}{lcccc}
			\hline
			\multicolumn{1}{l}{\bfseries Experiment} & \textbf{Units} &\multicolumn{1}{c}{\bfseries A} & \multicolumn{1}{c}{\bfseries B} & \multicolumn{1}{c}{\bfseries C} \\ \hline
			\hline
			Flow velocity & $\mathrm{cm/s}$ & 12.3 & 22 & 31 \\
			Average water column depth & $\mathrm{cm}$ & 7 & 14 & 12 \\
			Bedform average wavelength & $\mathrm{cm}$ & 15 & 20 & 37 \\
			Bedform average height & $\mathrm{cm}$ & 2 & 2.4 & 1.8 \\
			Side margin & $\mathrm{cm}$ & 1 & 1 & 1 \\
			Top margin & $\mathrm{cm}$ & 0.5 & 1 & 0.5 \\
			\hline
			Photo time step & $\mathrm{min}$ & 0.5 & 1 & 1 \\
			Photo interval & -  & 8 & 2 & 1 \\
			Bedform shaping & & Manual & Natural & Natural \\
			Reference & & \cite{Fox2014} & NA & \cite{Teitelbaum2021}\\
			\hline
		\end{tabular}
		\caption{Experimental and image processing parameters. The terms \textit{side margin} and \textit{top margin} refer to the margins of the domain within which the dye front was not considered due to data quality issues. The term \textit{photo interval} refers to the gap (in photo time steps) between the two photos used in optimization.}
		\label{table_exp_img_params}
	\end{table}

\section{Head and flux inference from dye front motion}
\label{sec: head-flux}

\subsection{Basic approach}

Dye front motion is linked to head distribution $\phi$ on the SWI by an indirect process. First, a boundary value problem (BVP) is solved, by which the steady-state groundwater flow equation and $\phi$ jointly determine that head distribution, $h(\bm x)$ throughout the domain. If $x$ is the horizontal coordinate that increases in the direction of flow, $z$ is the vertical coordinate, $x_\mathrm{min}$ and $x_\mathrm{max}$ are the upgradient and downgradient boundaries of the observed domain, $\kappa$ is a positive constant representing a characteristic influence depth of the HEF, and $\gamma(x)$ is understood as the elevation of the SWI, we formulate the BVP as:
\begin{align}
	\nabla^2 h &= 0, \label{eq: laplace}\\
	h(x,\gamma(x)), &= \phi(x)\\
	\frac{\partial h}{\partial z}(x, 0) &= 0, \\
	h(x_\mathrm{min},z) &= \phi(x_\mathrm{min})\exp\left[\kappa (z-\gamma(x_\mathrm{min}))\right],\\
	h(x_\mathrm{max},z) &= \phi(x_\mathrm{max})\exp\left[\kappa (z-\gamma(x_\mathrm{min}))\right].\label{eq: max BC}
\end{align}
We implicitly define $h=0$ to represent hydrostatic conditions, which we expect to obtain at depth, and we define the upgradient and downgradient boundary conditions to encode this sort of behavior.

By computation of $\nabla h$ and some correction terms, and application of Darcy's law at the dye front at time step $t$, the location of the dye front at a subsequent time step $t+n$ can be predicted, for $n \geq 1$. Conceptually, it suffices to compare the dye front at time step $t$ to that at the immediately subsequent time step $t+1$. In practice, however, sometimes it is necessary to compare against a slightly later timestep in order to get a clearer signal of the evolution of the dye front. For moving bedforms $n$ must be adjusted with care because the head boundary condition is continuously evolving, and therefore it is desirable to use $n$ as close to 1 as possible. We refer to $n$ as the "photo interval", and the values that we used are provided in Table \ref{table_exp_img_params}. An objective function is defined that penalizes the mismatch between the predicted dye front and the one actually observed at time step $t+1$, and contains a number of additional Tikhonov regularization terms. The objective function is minimized by iterative optimization of the function $\phi$ by gradient descent.

\subsection{Objective function}
	It is convenient to use set notation to formalize our ideas. we define $\Omega_i$ to be the dye-saturated subsurface region at time step $t$, which has boundary $\partial \Omega_t$. We define $\Gamma_i$ to be the portion of $\partial \Omega_t$ that lies on the the SWI, and define $\bm{x}$ as an arbitrary point $\langle x,z\rangle$. We use the symbol $\hat{\Omega}_{\phi,t}$ to represent the predicted dye-saturated region at step $t$, given $\phi$.
	
	As indicated, the primary goal of optimization is to determine the specified-head boundary condition, $\phi(x)$ and/or the corresponding Darcy flux distribution by minimizing the mismatch of predicted and observed plume evolution between a pair of photos that are close together in time. However, there are three additional constraints it is useful to apply. First, because only $\nabla h$ is observable, the distribution $\phi(x)$ is only identifiable up to a constant. To respect our definition that $h=0$ under hydrostatic conditions, absent flow induced HEF, we wish to enforce the constraint that the average value of $\phi(x)$ is zero. Second, the inverse problem is ill-posed, in that the effects of high-frequency oscillations in $\phi(x)$ decay with depth, and have limited or no observable impact on $\nabla h\vert_{\partial \Omega_t}$. Thus, a smoothing term that penalizes total variation of $\phi(x)$ is regularizing. Third, because dye is absent in the upwelling regions (where HEF is directed out of the bed), dependence of $\hat{\Omega}_t$ on  $\phi(x)$ is weaker there, and optimization may fit an unduly flat $\phi(x)$ in those regions to satisfy other regularization constraints, in effect underestimating the upwelling HEF. To counteract this, we can add another term to the objective function that penalizes imbalance between inward-directed and outward-directed HEF (such imbalance is allowed by BVP (\ref{eq: laplace}-\ref{eq: max BC}) because the upgradient and downgradient boundaries are open). We define the operator
	\begin{equation}
		A\{\Omega_t,\hat{\Omega}_{\phi,t}\}\equiv \vert(\Omega_t \cup \hat{\Omega}_{\phi,t}) \backslash (\Omega_t \cap \hat{\Omega}_{\phi,t})\vert/\vert \Omega_t \vert.
	\end{equation} 
	This is understood as representing the area that is in only one of $\Omega_i$ and $\hat{\Omega}_{\phi,t}$, normalized by the dye saturated area, which reduces as the regions become more similar and is zero iff they are identical.
	
	Armed with this definition, we formally define our objective functional
	\begin{equation}
		L\{\phi\} \equiv A\{\Omega_t,\hat{\Omega}_{\phi,t}\} + c_1 \left\vert\int_{\Gamma_i} \phi(\bm x) d \bm{x}\right\vert + c_2 \int_{\Gamma_i} \left\vert\frac{d \phi}{d \bm x}\right\vert d \bm{x} + c_3 \left\vert\int_{\Gamma_i} \bm{\nabla} h \cdot \bm{n}\ d\bm{x}\right\vert,
		\label{eq: objective}
	\end{equation}
	where $c_1, c_2, c_3$ are scalar weights. Recall that $h$ is implicitly a function of $\phi$ via the solution of the BVP (\ref{eq: laplace}-\ref{eq: max BC}). The optimal $\phi$ is the function that minimizes $L$. 

\subsection{Relating dye front velocity to pore water velocity}

	In the presence of diffusion and dispersion, the local velocity of the dye front (which is an arbitrary level set of solute concentration below which it becomes invisible to the photographic capture setup) does not generally move at the local pore water velocity. Consider a section of the dye front that can be modeled as semi-circular with local radius of curvature $R$. Assuming local radial symmetry, we adopt a coordinate system with its origin at a point of interest, $\bm x$, on the front, such that unit vector $\bm{\hat{r}}$ is normal to the dye front, pointing away from the center of curvature, and unit vector $\bm{\hat{s}}$ is orthogonal to $\bm{\hat{r}}$. Neglecting pore-scale transverse dispersion, we may write the advection-dispersion equation at a point on the front, noting that all derivatives in the direction of the level set are zero:
	\begin{equation}
		\frac{\partial c}{\partial t} = - v_r \frac{\partial c}{\partial r} + \frac{D_r}{R}\frac{\partial c}{\partial r} - v_s \cancelto{0}{\frac{\partial c}{\partial s}} + D_r\frac{\partial^2 c}{\partial r^2} + D_s\cancelto{0}{\frac{\partial^2 c}{\partial s^2}}.
		\label{eq: ade radial front}
	\end{equation}
	Here, the $v_r$ and $v_s$ are respectively the components of velocity in the $\bm{\hat{r}}$ and $\bm{\hat{s}}$ directions, and $D_r = \alpha_l v_r + D^*$, where $\alpha_l$ is the pore-scale longitudinal dispersivity and $D^*$ is the effective Fickian diffusivity of the dye. We propagate the instantaneous motion of point $\bm x$ on the dye front by means of a Taylor series expansion:
	\begin{equation}
		c(\bm{x}+\Delta_r\bm{\hat{r}},\Delta_t) \approx c(\bm{x},0) + \Delta_r\frac{\partial c}{\partial r} + \Delta_t\frac{\partial c}{\partial t}
		\label{eq: taylor}
	\end{equation}
	We \textit{define} $\Delta_r$ as the motion of the level set in time $\Delta_r$. thus, $c(\bm{x}+\Delta_r\bm{\hat{r}},\Delta_t) = c(\bm{x},0)$. We define $v_r' = \underset{\Delta_t\rightarrow 0}{\mathrm lim} \Delta_r/\Delta_t$ as the local velocity of the dye front. Then it follows from \eqref{eq: taylor} that 
	\begin{equation}
		v_r' = -\frac{\partial c}{\partial t}\left(\frac{\partial c}{\partial r}\right)^{-1}.
	\end{equation}
	Substituting in \eqref{eq: ade radial front} yields
	\begin{equation}
		v_r' = -\left(- v_r \frac{\partial c}{\partial r} + \frac{D_r}{R}\frac{\partial c}{\partial r} + D_r\frac{\partial^2 c}{\partial r^2}\right)\left(\frac{\partial c}{\partial r}\right)^{-1},
	\end{equation}
	which simplifies to
	\begin{equation}
		v_r' = v_r \underbrace{- \frac{D_r}{R}}_\text{RC} \underbrace{+ D_r\frac{\partial^2 c}{\partial r^2}\left(\frac{\partial c}{\partial r}\right)^{-1}}_\text{RD},
		\label{eq: velocity adjustment}
	\end{equation}
	Thus, the local velocity of the dye front relates to the local pore water velocity by the addition of two correction terms, RC, which stands for \textit{radius of curvature} and goes to zero for a planar front, and RD, which stands for \textit{ratio of derivatives}. The RD term represents the relative motion of the concentration level set representing the dye front relative to the level set corresponding to the line of inflection of the concentration near the leading edge of the plume where $\partial^2 c / \partial r^2 = 0$.

\section{Computer implementation}
\label{sec: implementation}
\subsection{Numerical approximation of terms}
	On the computer, the function $\phi(x)$ defining the head distribution on the SWI is represented by a vector, $\bm \phi$, of discrete values corresponding to locations along the $x$ axis separated by uniform spacing $\delta$, which are linearly interpolated. Similarly, the dye front $\partial \Omega_t$ is represented as a vector of points $\bm \omega$, where successive points are understood to be joined by straight line segments, and similarly with $\bm{\hat{\omega}}$ for the predicted dye front, $\partial \hat{\Omega}_t$. The set operator $A$ is approximated by a numerical operator $A'$ that concatenates $\bm{\omega}$ and $\bm{\hat{\omega}}$, treats them as the nodes of a (generally self-intersecting) polygon, and computes its area. We can express the objective function $L'$ that is actually minimized by the computer as 
	\begin{equation}
		L' = A' + \lvert\sum_i \bm{\phi}_i\rvert  + 0.05 \sum_{i>1} \lvert\bm{\phi}_i-\bm{\phi}_{i-1}\rvert.
	\end{equation}
	This approximates \eqref{eq: objective}, where $c_1=1$, $c_2=0.05\delta$, and $c_3=0$. 
	
\begin{figure}
		\centering
		\includegraphics[width=0.8\linewidth]{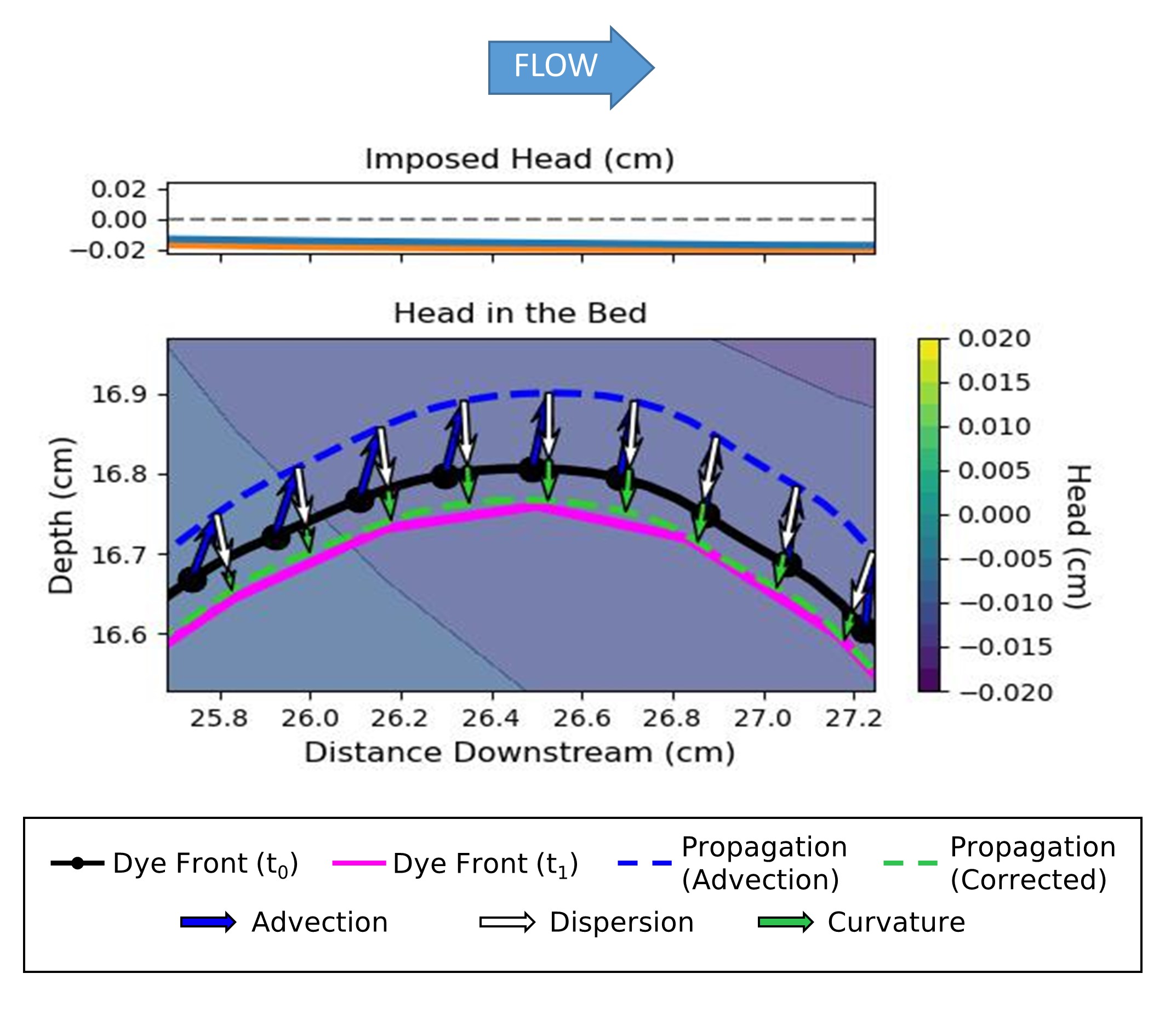}
		\caption{The components of dye front displacement for several nodes on the front in an upwelling region of the simulation discussed in Section \ref{sec: validation}. The simulated dye front is shown at two consecutive instants in time, $t_0$ and $t_1$, overlaid with the dye front predicted by propagating the nodes of the $t_0$ dye front with \eqref{eq: langevin}. For each node, the advection, RC (labeled \textit{curvature}), and RD (labeled \textit{dispersion}) vectors are shown. Note that the propagated front using advection alone moves in the wrong direction.}
		\label{fig:corrections}
	\end{figure}

	The motion of node $n$ of the dye front at time step $t$ can be approximated by multiplying \eqref{eq: velocity adjustment} by time step $\Delta_t$ and applying Darcy's law to determine local pore water velocity:
	\begin{equation}
		\hat{\bm{\omega}}_{t+1}^n = \bm{\omega}_{t}^n + \frac{k}{\theta}\frac{d h}{d \bm{r}}\Delta_t + \bm{k}_\mathrm{RC} + \bm{k}_\mathrm{RD},
		\label{eq: langevin}
	\end{equation} 
	where $k$ is hydraulic conductivity, and $\theta$ is porosity. Both of the correction vectors, $\bm{k}_\mathrm{RC}$ and $\bm{k}_\mathrm{RD}$ respectively represent the product of $\Delta_t$ with the RC and RD terms, must be estimated. The net effect of advection and the two correction vectors in propagating the dye front is shown in Figure \ref{fig:corrections}. All vectors must be considered, as they are often of comparable magnitude and, particularly in upwelling regions, the net motion of the dye front may be opposite to the direction of pore water flow. 
	
\begin{figure}
		\centering
		\includegraphics[width=.75\linewidth]{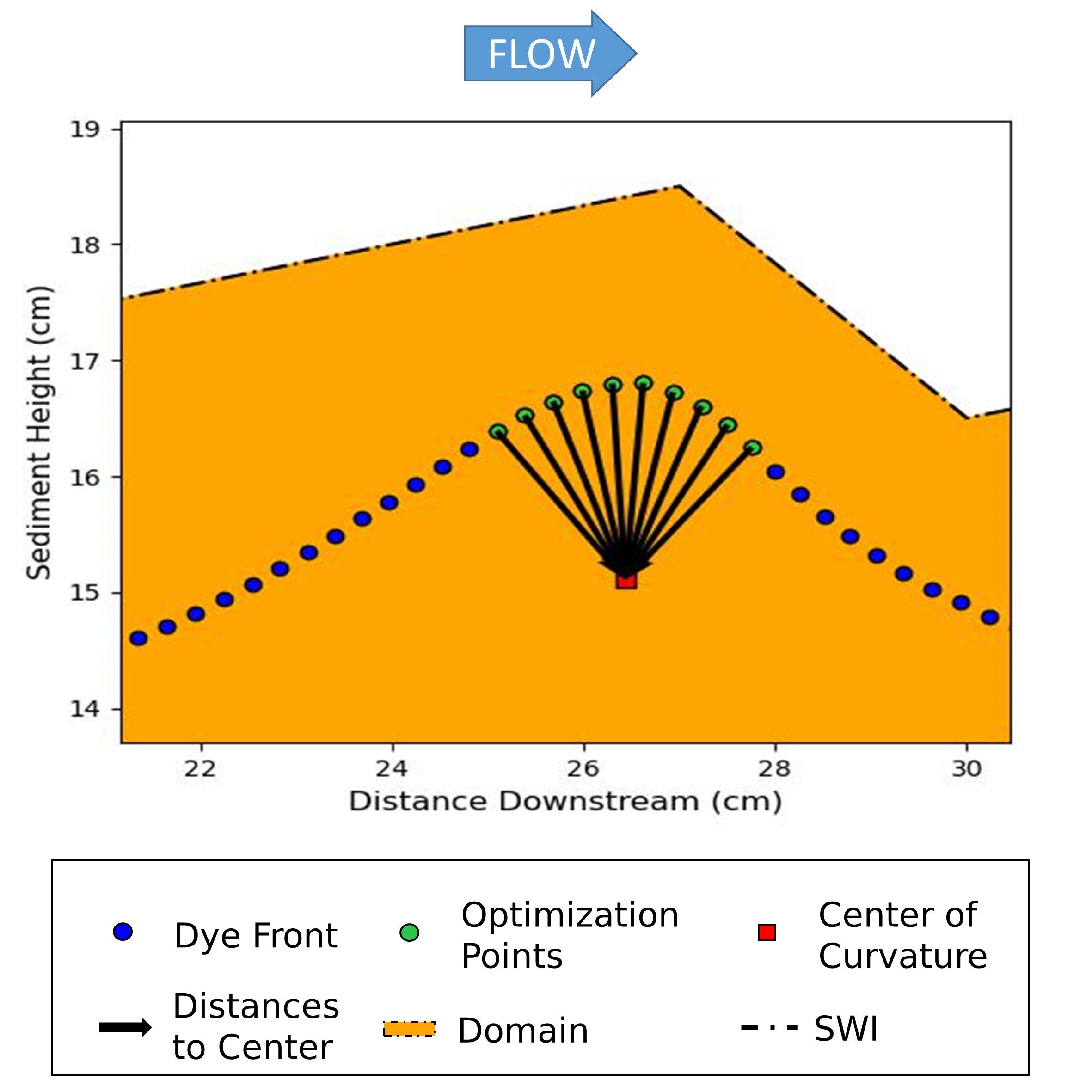}
		\caption{Finding the local radius of curvature, $R$. To find the $R$ at a point $\bm p$ along the dye front, the center of curvature $\bm c$ at location $\bm p$ is found by optimization. $\bm c$ is the point that is as close as possible to the same distance from $p$ ad its neighboring points on the discretized dye front, $\bm \omega_{t}$.}
		\label{fig:roptimization}
	\end{figure}

	For a given point on on dye front $\partial\Omega_t$,  $\bm{\omega}_{t}^n$ The RC term, $\bm{k}_\mathrm{RC}$, the dye-front-orthogonal unit vector, $\bm{\hat{r}}$, and the local radius of curvature, $R$, are all estimated by considering a subset of neighboring points $\bm{\zeta}^n_t$ in $\bm{\omega}_{t}$. The points are imagined to lie approximately on a circle, whose center, $\bm{c}(\bm{\omega}_{t}^n)$ is computed by finding the point that minimizes the variance of the distances between $\bm c$ and the considered points on the dye front, as illustrated in Figure \ref{fig:roptimization}). In general, the number of neighbor points used to compute $R$ depends on the photography equipment and lighting in use during the dye tracer test, as well as on the resolution of discretization of $\partial \Omega$. Formally, 
	\begin{equation}
		\bm{c}(\bm{\omega}_{t}^n) \equiv \underset{\bm{x} \in \mathbb{R}^2, \bm{p} \in \bm{\zeta}^n_t}{\min}  \mathrm{Var}\ \lVert \bm{x} - \bm{p} \rVert_2.
	\end{equation}
	 Using this point, it is then possible to compute
	\begin{align}
		R(\bm{\omega}_{t}^n) &= \lVert \bm{\omega}_{t}^n - \bm{c}(\bm{\omega}_{t}^n)  \rVert_2, \\
		\bm{\hat{r}}(\bm{\omega}_{t}^n)&= \frac{\bm{\omega}_{t}^n - \bm{c}(\bm{\omega}_{t}^n)}{R(\bm{\omega}_{t}^n)},\\
		\bm{k}_\textrm{RC}(\bm{\omega}_{t}^n) &= \frac{\alpha_l v_r + D^*}{R(\bm{\omega}_{t}^n)}\Delta_t.
	\end{align}
	Radius of curvature $R$ varies along the dye front, and larger $R$ corresponds to a smaller RC term. Where a point of inflection exists in the curvature of the dye front, there is a discontinuity in the direction of $\bm{\hat{r}}$, However, because $R \rightarrow \infty$ at the point of inflection, the effect of this discontinuity on $\bm{k}_\mathrm{RC}$. is negligible.
	
	If concentration information is available in the vicinity of the dye front, then the RD term can be estimated directly from evaluation of the first two derivatives of concentration:
	\begin{equation}
		 \bm{k}_\mathrm{RD} = \left.\frac{\partial^2 C}{\partial \bm{r}^2} \left(\frac{\partial C}{\partial \bm{r}}\right)^{-1}\right\vert_{\bm{x} = \bm{\omega}_{t}^n} \bm{\hat{r}}(\bm{x}).
	\end{equation}
	Where concentration information is not available, an ansatz must be specified for the RD term and coefficients must be jointly fit with those of $\bm{\phi}_t$. By running advective-dispersive transport simulations (discussed in detail in Section \ref{sec: validation}) to generate the synthetic data and concentration contours, we found that RD generally exhibits a strong spatial pattern, increasing continuously from its minimum value in the center of the upwelling regions, to its maximum magnitude near the center of the downwelling regions. Introducing symbol $x_m$ denote the locally smallest-$z$ location on the dye front, nearest to point $\bm{x}$, we introduce the ansatz
	\begin{equation*}
		\bm{k}_\mathrm{RD} =  \left[a(x-(x_m+c))^2 + b\right]\bm{\hat{r}}(\bm{x}),
		\label{eq: ansatz}
	\end{equation*}
	where $a$, $b$, and $c$ are free parameters.
	
\subsection{Optimization procedure}
	\label{subsec:optimizationprocedure}
	
	For a given time step, $t$, the discretized head boundary condition vector, $\bm{\phi}_t$ is initially set to a small-magnitude, zero-mean random vector by independently sampling each of its entries from distribution $U\left( -10^{-6}, 10^{-6}\right) $. The pore water flow field resulting from this initial guess is calculated via finite volume evaluation of (\ref{eq: laplace}-\ref{eq: max BC}) and the nodes of the discretized observed dye front $\bm{\omega}$ are then propagated using \eqref{eq: langevin}, generating a predicted dye front $\partial \hat{\Omega}_{t+n}$ for time step $t+n$ (Figure \ref{fig:optimizationpertimestepwithimages}c). $\partial \hat{\Omega}_{t+1}$ is then compared to the dye front $\partial \Omega_{t+n}$ observed at step $t+n$, by evaluating operator $A'$. The evaluation is accomplished by generating a closed polygon from the two vectors of nodes and calculating its area using the geometric analysis python library Shapely \citep{Gillies2007}. Optimization is accomplished by iterative adjustment of $\bm{\phi}_t$ as well as $a$, $b$, and $c$ in \eqref{eq: ansatz} using the \texttt{minimize} function with the bounded BFGS algorithm, L-BFGS in SciPy \citep{2020SciPy-NMeth}.

\subsection{Handling data set limitations}   
	Head identification operates conceptually by inversion from the velocities along the dye front. Because the RC and RD correction factors are functionally nuisance parameters that interfere with the ability to infer velocities at the dye front from local motion of the front, it is easiest to solve the inverse problem where the magnitudes of the RD and RC vectors are small relative to the magnitude of the advection vector. In our experience, the strongest advective signal occurs at early-to-intermediate times in the dye tracer test. At very early time, immediately after dye injection, concentration gradients are strong, contributing to dominance of the RD term. At late time, the dye front reaches relatively deep below the SWI, where advective velocities are lower. At this point, the RC term and RD term (this time due to diffusion dominance) grow in respective relevance. If only dye front location is available, we recommend usage of early/intermediate times for most straightforward inversion. We stress that this is not a theoretical limitation, simply a limitation of our experimental data set. Both the early- and late-time limitations originate from the availability of, in effect, only one contour line of constant dye concentration. This necessitates specification of an ansatz and joint inversion to recover RD simultaneously with the head distribution. RC is directly observable, but is sensitive to inferred front shape. By measuring concentration throughout the dye plume with a more sensitive optical apparatus or mechanically, as in \cite{Janssen2012}, RD becomes a directly observable quantity and RC identification accuracy is improved.

\subsection{Validation}
\label{sec: validation}

	Two forms of validation were employed: (a) Experiment A was run at a sufficiently low flow rate that bedforms remained stationary throughout, implying that the head distribution recovered must be the same at each time step, which was checked. (b) A completely synthetic test was run in which contours were generated by numerical solution of the advection-dispersion equation with a specified boundary condition, and the inversion was run to verify it recovered the specified boundary condition. The two validations have different strengths. Validation (a) only checks consistency, not correctness, but tests all parts of the computational chain, including photographic feature identification, and uses realistic parameters. Validation (b) checks directly for correctness because the result is known, at the cost of using artificial data and nonphysically high diffusivity to overcome the effects of numerical dispersion. 
	
	For validation (a), the results of recovering the head boundary condition at distinct time steps are virtually identical, regardless of the changed position of the dye plume (Figure \ref{fig:optresultonimagest31t32zoomedoutfromppt}). We now discuss validation (b) in detail.

	\subsubsection{Generation of synthetic data}
	\label{subsec:gen-of-synth-data}
	
	A synthetic flow field was generated employing (\ref{eq: laplace}-\ref{eq: max BC}) as instantiated in the flow model described in \cite{Teitelbaum2022}. Then, a finite volume solver was used to simulate a dye tracer test with similar conditions to Experiment A. Because numerical dispersion is significant but its magnitude cannot be predicted \textit{a priori} when using an Eulerian solver in the presence of advection, a sufficiently large diffusivity was selected to dominate numerical (and pore-scale) dispersion---otherwise model error owing to the un-modeled dispersion would interfere with our ability invert the model. Dye propagation was thus simulated via the advection-diffusion equation,
	\begin{equation}
		\label{eq:ade}
		\frac{\partial c}{\partial t} = -\bm{v}\cdot\nabla c + D^* \nabla^2 c
	\end{equation}
	on the same domain used to solve for the head distribution, $h(\bm x)$ using equations (\ref{eq: laplace}-\ref{eq: max BC}).  and the velocity was computed
	\begin{equation}
		\bm{v} = \frac{k}{\theta}\nabla h.
	\end{equation}
	
	The modeled sediment bed was 60 cm long with a sediment depth of 16.5 cm (Figure \ref{fig:optvsimposedzoomedout}). The top boundary of the domain was shaped like four triangular bedforms of length 15 cm each and height 2 cm, with bedform crest occurring 12 cm downstream of its preceding trough. A sinusoidal head distribution with period equal to 15 cm and amplitude 0.02 cm was imposed. The finite volume grid was rectilinear, with a horizontal grid resolution of 0.33 cm, and a vertical grid resolution of 0.17 cm. Intrinsic model parameters for this simulation are given in Table \ref{table_synth_params}. The first dye front $\partial \Omega_{t}$ was created by allowing the model to run for a simulated duration of 15.5 minutes, which corresponded to the earliest analyzed time from Experiment A. Each time step represented a simulated duration of 30 seconds, the same as the time between photos from Experiment A. The subsequent dye front $\partial \Omega_{t+n}$ was created by allowing the model to run for two more time steps, i.e. "photo interval" $n$ = 2 in this case. Propagation of a conservative tracer corresponding to the dye was simulated using the ADE (\ref{eq:ade}). Boundary dye concentration at the surface was prescribed as 0.088 g/L. 
	
	\begin{center}
		\begin{tabular}{lccc}
			\hline
			\textbf{Description} & \textbf{Units} & \textbf{Symbol} & \textbf{Value} \\
			\hline \hline
			Porosity & - & $\theta$ & $0.33$ \\
			Diffusivity & $\mathrm{cm^2 / s}$ & $D^*$ & $1.33 \cdot 10^{-3}$ \\
			Hydraulic conductivity	& $\mathrm{cm/s}$  & $k$  & $0.12$ \\
			\hline
			
		\end{tabular}
		\captionof{table}{Model parameters for generation of synthetic data set.}
		\label{table_synth_params}
	\end{center}
	
	\subsubsection{Optimization on synthetic data}
	\label{subsec:opt-on-synth-data}
	
\begin{figure}[h]
		\centering
		\includegraphics[width=0.9\linewidth]{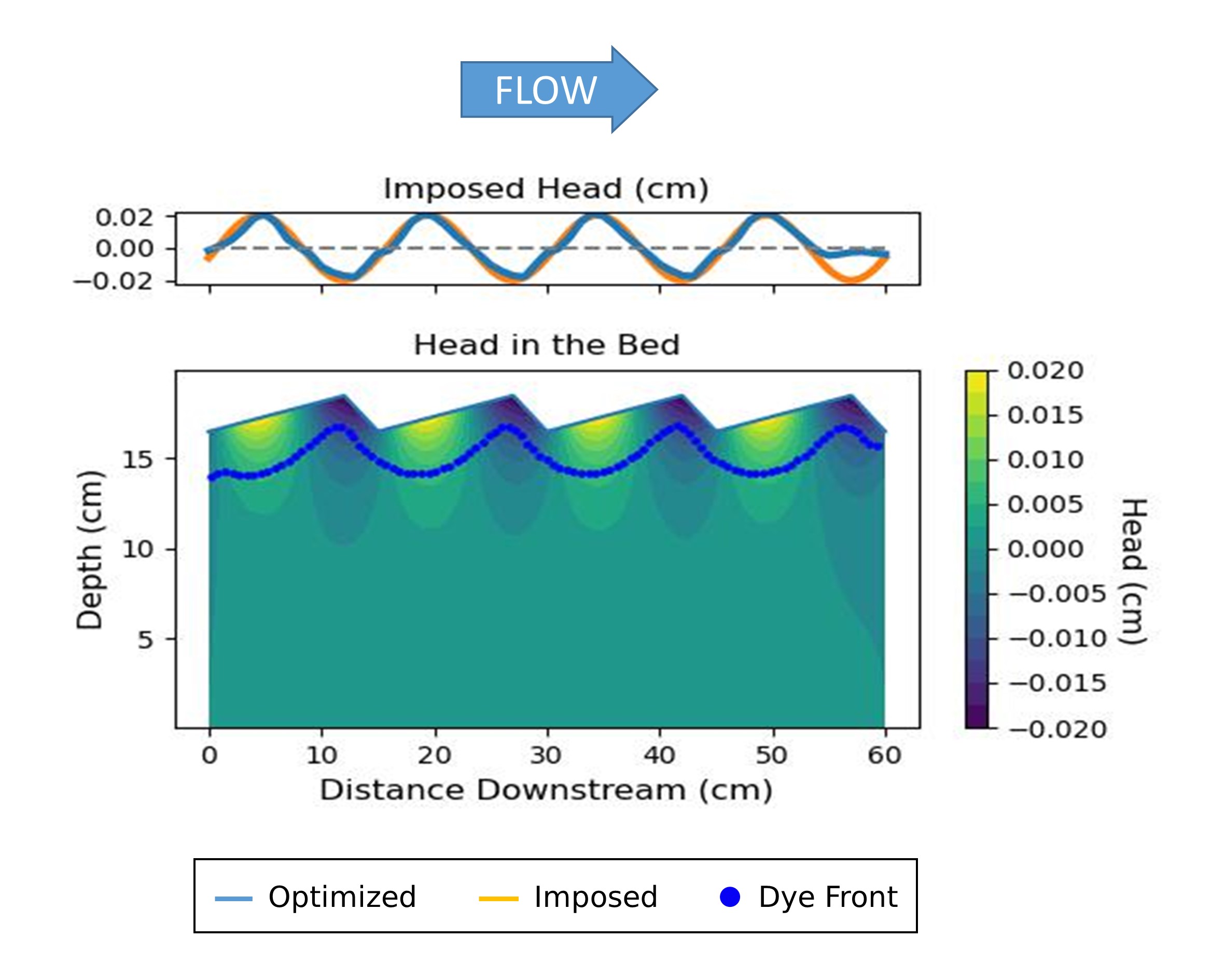}
		\caption{Result of optimization to recover the head boundary condition on the synthetic data set. Approximate head distribution identified by optimization (blue line) and the true, imposed sinusoidal distribution (orange line) are presented on the top axes at one instant in time. A simulated dye front is also shown overlaid on the spatial head distribution in the domain on the lower axes.}
		\label{fig:optvsimposedzoomedout}
	\end{figure}

	The head boundary condition obtained by optimizing on the synthetic data closely resembled the one that was imposed in generating the data (Figure \ref{fig:optvsimposedzoomedout}). Both the magnitudes and locations of head maxima were very close between the optimized and imposed curves. The values of local head minima were slightly less negative than those of the imposed curve, and their horizontal position was slightly downstream of those of the imposed curve. The optimized curve flattened somewhat relative to the imposed one where both curves crossed zero in the increasing direction. The greatest difference between the two curves occurred near the right boundary of the domain. In that region, the imposed curve exhibits a local minimum typical of a sinusoidal curve, while the optimized curve becomes almost flat, taking values close to zero.
	
\section{Results}
\label{sec:results}

\subsection{Bedform and dye front time series}

We captured multiple-hour data sets for three dye tracer tests performed under different flow conditions. The captured data include sediment bed profiles and dye fronts extracted by image processing from time lapse photos taken every 30-60 seconds. Experiment A featured the slowest flow velocity (0.12 m/s) and stationary hand-shaped triangular bedforms of length 15 cm and height 2 cm. Dye plumes form underneath each bedform in the classic conchoid shape. Experiment B was conducted at a flow speed of 0.22 m/s. The time lapse footage includes several bedforms having lengths of roughly 15-20 cm which change over the course of the tracer test, and average height 2.4 cm. The bedforms move relatively slowly, advancing less than one bedform wavelength over the course of 967 minutes of footage. The dye front still forms conchoid shapes which are recognizable even after distortion under the changing bed shape. Experiment C was performed at the fastest velocity, 0.31 m/s. As could be expected, the bedforms in this test were longer and less tall than in the slower-flow tests, having average length and height of 37 cm and 1.8 cm respectively. More generally, the shape of the sediment bed in this test often does not include a full set of defined bedforms, instead featuring a flat or gently sloping shape along much of the domain. The bed motion is less regular, and bedforms that do appear advance multiple wavelengths during the 612 minutes of footage. After about four hours, the initially conchoid dye plumes have merged to form one large dye plume.

For Experiments B and C, the optimization procedure was run on five consecutive pairs of images. Because the images were taken close together in time (see Table \ref{table_exp_img_params}), the domain and dye front exhibited minimal change over the course of the five images. The results of these five optimizations were then averaged together to produce an estimate of the head top boundary condition.

The data resulting from this work have all been archived at the HydroShare repository: \texttt{https://doi.org/10.4211/hs.5cbece3ef6ad4e08a08d611681e1f800}. Uploaded data include:
\begin{itemize}
	\item[\ding{227}] Digitized sediment profiles, extracted every 30-60 seconds from time lapse footage of the dye tracer tests.
	\item[\ding{227}] Digitized dye front locations, extracted every 30-60 seconds from time lapse footage of the dye tracer tests.
	\item[\ding{227}] Video files of each of the dye tracer tests, with the extracted bed profiles and dye fronts highlighted.
	\item[\ding{227}] Individual estimated head boundary conditions at the SWI for the various experiments.
\end{itemize}

In the following sections, we discuss head distribution recoveries for each of the three experiments. 

\subsection{Experiment A: lowest velocity, hand-formed bedforms}
\label{subsec:resultsstationarybedforms}

\begin{figure}
	\centering
	\includegraphics[width=1\linewidth]{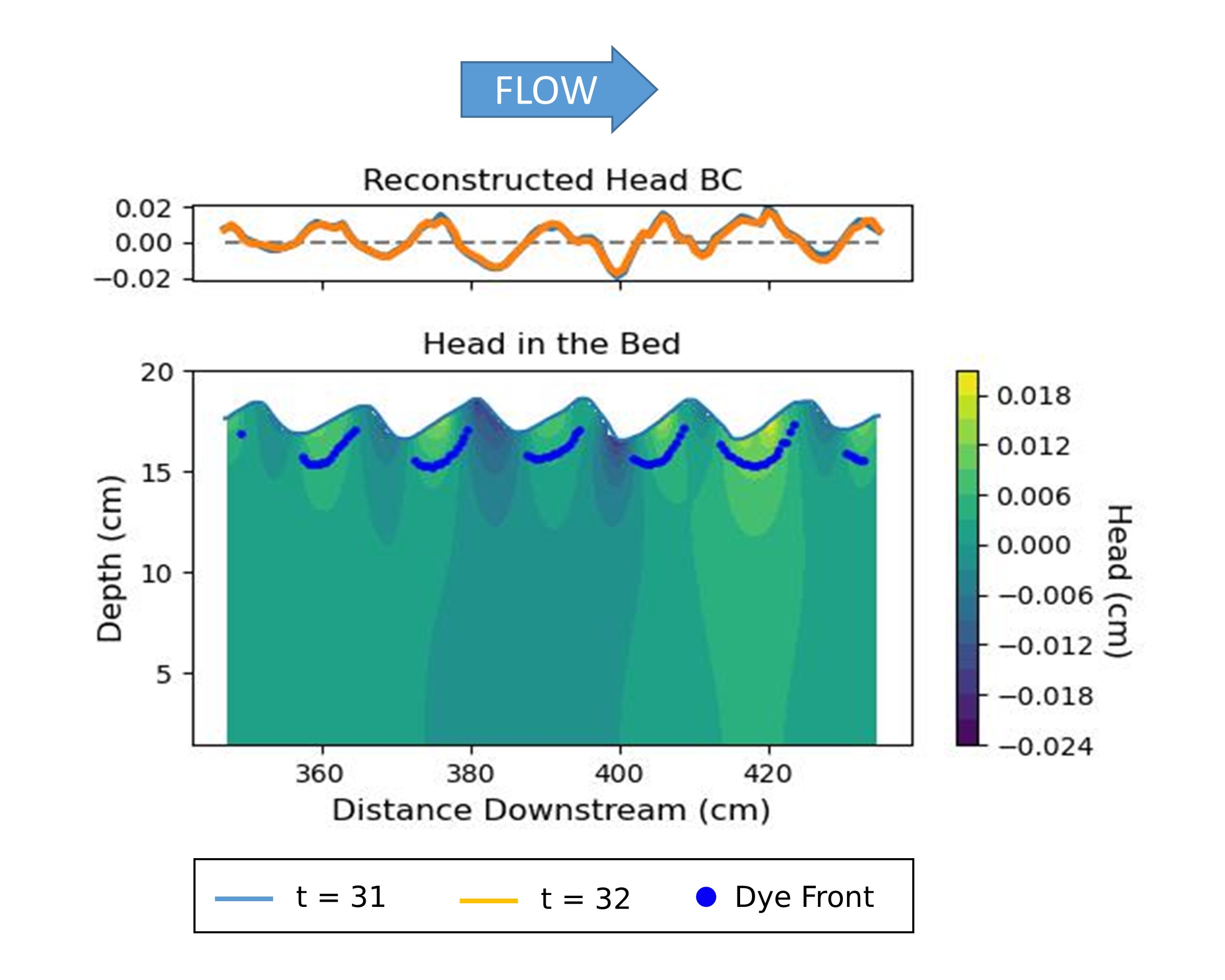}
	\caption{Results of optimizing for the head boundary condition at two consecutive times during the Experiment A dye tracer test are shown. (Results are so similar that the lines are indistinct.) The bottom axes show the head distribution in the domain corresponding to the recovered head BC, as well as the portions of the dye front that were used for recovery (portions too close to the SWI were not used for calibration).}
	\label{fig:optresultonimagest31t32zoomedoutfromppt}
\end{figure}

Figure \ref{fig:optresultonimagest31t32zoomedoutfromppt} shows the recovered SWI head distribution, which exhibits periodic behavior. Local head maxima and minima are located at the stoss and lee bedform faces, respectively. Thus the general shape is similar to the boundary condition suggested by \cite{Elliott1997}. The Elliott and Brooks half-amplitude $H_m$ corresponding to these experimental conditions is approximately 0.0233 cm. This is close to the global maximum (0.0226 cm) and minimum (-0.0213 cm) obtained for the head boundary condition in this procedure. The differences between local maxima and minima in our results are somewhat less than the Elliott and Brooks amplitude. 

Notably, despite the fact that the bedforms are all similarly shaped to each other, there is significant variation from bedform to bedform in the corresponding induced head distribution.

\begin{figure}
	\centering
	\includegraphics[width=0.8\linewidth]{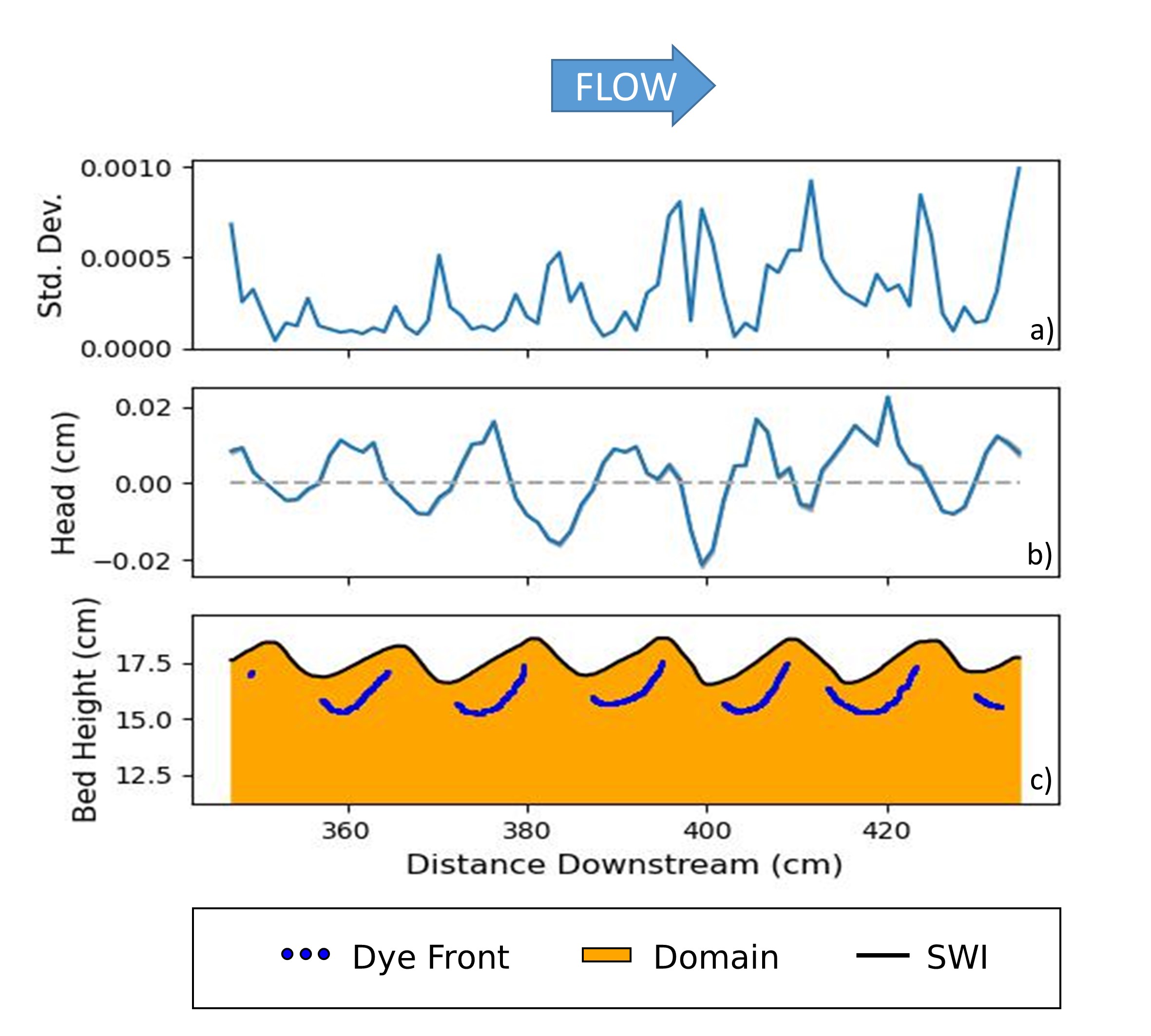}
	\caption{a) Standard deviation of head BC at interpolation points. b) Mean +/- standard deviation of head BC at interpolation points. c) Sediment bed with dye front data prepared as in Section \ref{subsec:dyetracertests}.}
	\label{fig:photomeanandstd}
\end{figure}

The optimization on the time series image was performed ten times in order to assess the uncertainty in the optimization procedure (Figure \ref{fig:photomeanandstd}). The maximal standard deviation along the domain was $9.9 \cdot 10^{-4}$. The maximal head BC magnitude along the domain was 0.0226 cm. The ratio between the maximal standard deviation and the maximal head magnitude was 0.044 cm, or less than 5\%. The maximal standard deviation occurred at the domain right boundary. Other local maxima occurred at the domain left boundary and in the vicinity of local minima in the head BC.

The instantaneous HEF influx along the domain was also calculated for each of the ten optimizations in the uncertainty analysis. The mean HEF influx was 0.48 $\mathrm{cm}^3$ for the 30 seconds represented by the analyzed photo. The standard deviation of the HEF influx was $3.73\cdot10^{-3}$. Thus, the coefficient of variation of the HEF influx was $7.76\cdot10^{-3}$, or less than 1\%.

As an additional check, the head boundary condition was estimated at every hour over the course of the dye test. The result was essentially unchanged in terms of the location of local head minima and maxima. However, the amplitude of the result appeared to become damped over time; the magnitude of local head gradients generally decreased for later times. In particular, local head minima were not as negative at later times.

\subsection{Experiment B: intermediate velocity, naturally-evolved bedforms}
\label{subsubsec:expb}

\begin{figure}[h]
	\centering
	\includegraphics[width=1\linewidth]{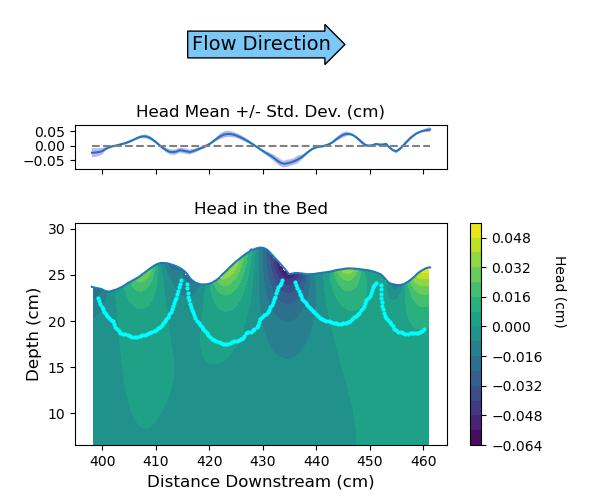}
	\caption{Optimization results from the Experiment B dye tracer test. Top panel shows mean and standard deviation of optimization results from individual time steps. Bottom panel shows the shape of the sediment bed at time t = 60, as well as the head distribution throughout the sediment bed resulting from the mean in the top panel. The cyan dots in the bottom panel show the shape of the dye front at time t = 60.}
	\label{fig:headmeanandstd2022t5862}
\end{figure}

The recovered head distribution for Experiment B is shown in Figure \ref{fig:headmeanandstd2022t5862}. The recovery was repeated five times using the photos that were taken between 58 and 62 minutes from the start of the dye tracer test, a period in time during which the SWI shape was relatively constant. This allowed comparison of recovery reliability by computation of the point-wise standard deviation of the recovered distributions, as in Experiment A. 

Distinct local maxima and minima still appear at the elevated flow rate in this experiment, and head maxima at the SWI generally scaled with bed height. The global head gradient (global head maximum minus global head minimum) for the estimated mean was 0.119 cm, which compares with half-amplitude for a sinusoidal BC as computed by \cite{Elliott1997} is 0.065 cm. Thus the relative error of the observed head gradient relative to the Elliott and Brooks head gradient was computed to be 9.8\%. In this case the global head gradient was calculated using the local head maximum near x = 423 and not the one near x = 460 as the latter is near the downstream domain boundary, which is a region of greater uncertainty, as discussed above.

The largest point-wise standard deviation occurred at the domain left boundary, and constituted 13.1\% of the global head gradient. The next largest standard deviation is 11.2\% at approximately x = 436.5, an upwelling zone. Elsewhere, the point-wise standard deviations of the recovered heads are below 10\% of global head gradient.

\subsection{Experiment C: highest velocity, naturally-evolved bedforms}
\label{subsubsec:expc}

\begin{figure}[h]
	\centering
	\includegraphics[width=1\linewidth]{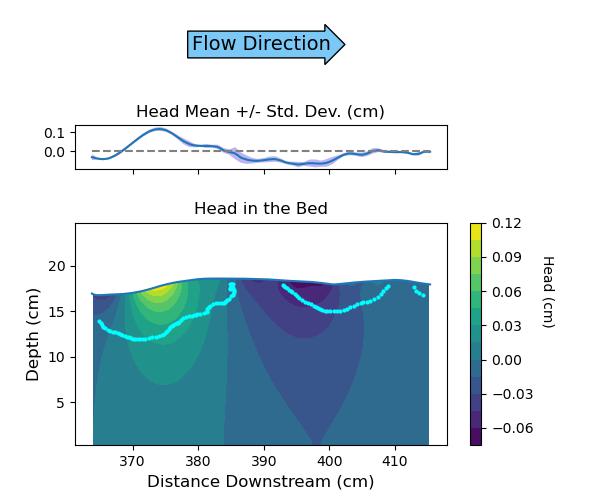}
	\caption{Results of optimizing on t = 28-32 minutes elapsed for the Experiment C tracer test. Top panel shows mean +/- standard deviation of the individual estimates of the head BC. Bottom panel shows the shape of the sediment bed at t = 30 and head throughout the bed in light of the mean head BC from the top panel. Cyan dots in the bottom panel show the dye front at time t = 30.}
	\label{fig:headmeanandstd2018t2832}
\end{figure}

Similar to Experiment B, in Experiment C, the five time steps that were analyzed (here from 28 to 32 minutes elapsed from the start of the tracer test) Results are shown in Figure \ref{fig:headmeanandstd2018t2832}. Much higher scour is observed at the high flow rates in this experiment, with relatively muted elevation changes in the SWI. Nevertheless, the local head maxima remain associated with the upward-sloping regions of the sediment bed (near x = 375 and x = 408), while the head minimum occurs along the downward-sloping region near x = 395. As with Experiment B, changes in head scaled with changes in bed height along the SWI. The global head gradient was 0.189 cm, compared with the Elliott and Brooks half-amplitude for these conditions of 0.1 cm, corresponding to a relative error of 6.7\%. 

Standard deviation of the estimated head BC was greatest near x = 385, amounting to 17.2\% of the global head gradient. The region of the next-greatest standard deviation was near x = 397, with a standard deviation of 11.2\% of the global head gradient. Both of these regions of relatively high uncertainty correspond to segments of the dye front that lie within the upwelling zone. Elsewhere, the standard deviation is less than 10\% of the global head gradient, as in Experiment B.

\section{Discussion}
\label{sec:discussion}

The method discussed here exhibits a high degree of precision. In Experiment A, bedforms were stationary and the true head boundary function, $\phi$ known a priori to be fixed over time. The point-wise (i.e., fixed location) variability of the recovered head distributions at various time steps was thus a proxy for the precision of the method. The point-wise standard deviation of the optimized $\hat{\phi}$ was less than 5\% of its RMS value at all points (Figure \ref{fig:photomeanandstd}), and frequently less. Local maxima of standard deviation in this experiment occurred at two sets of locations: at the side boundaries of the domain, and in the vicinity of local head minima. The downstream boundaries of the observed area exhibited higher uncertainty as dye front data relevant to constraint of the head BC was out of frame. The local head minima, meanwhile, correspond to upwelling zones where the dye front is absent. Additionally, the direction of pore water flow is approximately parallel to the dye front in these zones. These two factors combined mean that local change in the head BC has a relatively weak impact on dye front propagation. Thus the head BC in upwelling zones is less constrained by the dye front.

The optimized head BCs for Experiments B and C generally were in agreement with literature. Location of features of the head BC agree with what is known about head imposed by bedforms: local head maxima occurred at uphill regions of the SWI, and local head minima occurred at downhill regions of the SWI. The same is true for the scale of the head BC: global head gradients were in close agreement with the relationship formulated by \cite{Elliott1997}, and magnitude of head gradient scaled with magnitude of bed profile gradient.

The results shown here highlight the need for a method of measuring head distributions throughout an arbitrary, generally transient SWI. In the head boundary condition estimated at early time for Experiment A, the local head drops across bedforms varied between 0.019 and 0.031 cm, a variation of 38.8\%. This variation occurred on bedforms that were manually sculpted to be identical using a stencil, i.e., on bedforms that are similar beyond what can be found in nature. If such a great variation occurs on "identical" bedforms that do not change shape, a far greater degree of variation can be expected to occur under natural bedforms, especially under moving bed conditions. Bedform shape and progression downstream are often treated as processes that fluctuate in a regular manner about some mean. However, it has been shown that streambeds can become intermittently stationary \citep{Dallmann2021,Risse-Buhl2023}, which can lead to extended amounts of time spent in arbitrary formations. In order to better understand the resulting impact on HEF, a method of estimating head locally is necessary. Ultimately, the techniques of this paper enable production of experimental data sets linking SWI bedform shape and induced head, which will ultimately inform non-CFD predictive models.

\section{Summary and conclusions}
\label{sec:conclusion}

We have presented a new data set of time series of spatially heterogeneous bedform profiles as well as corresponding head distributions. To our knowledge, the data obtained using this method are the first for naturally-shaped sand bedforms. The head distributions were recovered via a new, optically-based calibration method for inverting dye tracer tests. This method is non-invasive, which allows it to be applied over bedforms of arbitrary and transient shape, unlike the approaches used in previous experimental work on the topic. 

The data sets we recovered have the potential to inform future non-CFD predictive models relating bedform profile to head distribution. The non-invasive inversion method is particularly suitable for estimating head at the SWI at various points in time during moving bed conditions, and the physically-constrained regularization constraints that we have developed have the potential to be employed in a wide range of analyses in which the hydrological and sedimentological aspects of bedform motion are studied in tandem, which has still largely not occurred.

A current limitation of the inversion approach is its reliance on a correction term (RD) that is not directly observable from experimental data and which must be jointly inverted with the quantities of interest. Usage of a tracer in which concentration variations adjacent to the front are directly observable, such as the one employed by \cite{Janssen2012}, will eliminate this difficulty. All of the mathematics we developed still apply in such a case.

A potentially valuable future application of this work would be to extend it to a variety of experimental conditions and sediment bed configurations in order to build a library of sediment bed profiles and corresponding SWI head distributions. This library could then be used to train a machine learning algorithm to predict head distribution solely based on bed shape.

\section*{Acknowledgments}
This work was supported by Deutsche Forschungsgemeinschaft (DFG) Research Grant LE 1356/7-1, National Science Foundation and United States-Israel Binational Science Foundation NSF-BSF joint program in Earth Sciences award EAR-1734300, and Israel Science Foundation Individual Research Grant 647/21. SKH holds the Helen Ungar Career Development Chair in Desert Hydrogeology.

\bibliographystyle{plainnat}
\bibliography{./References}

\begin{thebibliography}{47}
\providecommand{\natexlab}[1]{#1}
\providecommand{\url}[1]{\texttt{#1}}
\expandafter\ifx\csname urlstyle\endcsname\relax
  \providecommand{\doi}[1]{doi: #1}\else
  \providecommand{\doi}{doi: \begingroup \urlstyle{rm}\Url}\fi

\bibitem[Ashley(1990)]{Ashley1990}
Gail~M. Ashley.
\newblock {Classification of Large-Scale Subaqueous Bedforms: A New Look at an
  Old Problem}.
\newblock \emph{SEPM Journal of Sedimentary Research}, 60\penalty0
  (1):\penalty0 160--172, 1990.
\newblock ISSN 1527-1404.
\newblock \doi{10.1306/212f9138-2b24-11d7-8648000102c1865d}.

\bibitem[Bencala and Walters(1983)]{Bencala1983}
Kenneth~E Bencala and Roy~A Walters.
\newblock {Simulation of solute transport in a mountain pool-and-riffle stream:
  A transient storage model}.
\newblock \emph{Water Resources Research}, 19\penalty0 (3):\penalty0 718--724,
  1983.
\newblock \doi{https://doi.org/10.1029/WR019i003p00718}.
\newblock URL
  \url{https://agupubs.onlinelibrary.wiley.com/doi/abs/10.1029/WR019i003p00718}.

\bibitem[Bennett and Best(1995)]{Bennett1995}
S.~J. Bennett and J.~L. Best.
\newblock {Mean flow and turbulence structure over fixed, two‐dimensional
  dunes: implications for sediment transport and bedform stability}.
\newblock \emph{Sedimentology}, 42\penalty0 (3):\penalty0 491--513, 1995.
\newblock ISSN 13653091.
\newblock \doi{10.1111/j.1365-3091.1995.tb00386.x}.

\bibitem[Boano et~al.(2014)Boano, Harvey, Marion, Packman, Revelli, Ridolfi,
  and W{\"{o}}rman]{Boano2014}
F.~Boano, J.~W. Harvey, A.~Marion, A.~I. Packman, R.~Revelli, L.~Ridolfi, and
  A.~W{\"{o}}rman.
\newblock {Hyporheic flow and transport processes: Mechanisms, models, and
  biogeochemical implications}.
\newblock \emph{Reviews of Geophysics}, 52\penalty0 (4):\penalty0 603--679,
  2014.
\newblock ISSN 0096-3941.
\newblock \doi{10.1002/2012RG000417}.
\newblock URL \url{http://doi.wiley.com/10.1002/2012RG000417}.

\bibitem[Bradski(2000)]{Bradski2000}
G.~Bradski.
\newblock {The OpenCV Library}.
\newblock \emph{Dr. Dobb's Journal of Software Tools}, 2000.

\bibitem[Cardenas(2008)]{Cardenas2008}
M.~Bayani Cardenas.
\newblock {Surface water-groundwater interface geomorphology leads to scaling
  of residence times}.
\newblock \emph{Geophysical Research Letters}, 35\penalty0 (8):\penalty0 1--5,
  2008.
\newblock ISSN 00948276.
\newblock \doi{10.1029/2008GL033753}.

\bibitem[Cardenas(2009)]{Cardenas2009}
M~Bayani Cardenas.
\newblock {Stream-aquifer interactions and hyporheic exchange in gaining and
  losing sinuous streams}.
\newblock \emph{Water Resources Research}, 45\penalty0 (June):\penalty0 1--13,
  2009.
\newblock \doi{10.1029/2008WR007651}.

\bibitem[Cardenas and Wilson(2007)]{Cardenas2007b}
M.~Bayani Cardenas and John~L. Wilson.
\newblock {Dunes, turbulent eddies, and interfacial exchange with permeable
  sediments}.
\newblock \emph{Water Resources Research}, 43\penalty0 (8):\penalty0 1--16,
  2007.
\newblock ISSN 00431397.
\newblock \doi{10.1029/2006WR005787}.

\bibitem[Chen et~al.(2015)Chen, Cardenas, and Chen]{Chen2015}
Xiaobing Chen, M.~Bayani Cardenas, and Li~Chen.
\newblock {Three-dimensional versus two-dimensional bed form-induced hyporheic
  exchange}.
\newblock \emph{Water Resources Research}, 51\penalty0 (4), 2015.
\newblock ISSN 1093-474X.
\newblock \doi{10.1111/j.1752-1688.1969.tb04897.x}.

\bibitem[Chen et~al.(2018)Chen, Cardenas, and Chen]{Chen2018}
Xiaobing Chen, M.~Bayani Cardenas, and Li~Chen.
\newblock {Hyporheic Exchange Driven by Three-Dimensional Sandy Bed Forms:
  Sensitivity to and Prediction from Bed Form Geometry}.
\newblock \emph{Water Resources Research}, 54\penalty0 (6):\penalty0
  4131--4149, 2018.
\newblock ISSN 19447973.
\newblock \doi{10.1029/2018WR022663}.

\bibitem[Dallmann et~al.(2021)Dallmann, Phillips, Teitelbaum, Cifuentes, Sund,
  Schumer, Arnon, and Packman]{Dallmann2021}
J.~Dallmann, C.~B. Phillips, Y.~Teitelbaum, Edwin Y.~Saavedra Cifuentes,
  N.~Sund, R.~Schumer, S.~Arnon, and A.~I. Packman.
\newblock {Bedform segregation and locking increase storage of natural and
  synthetic particles in rivers}.
\newblock \emph{Nature Communications}, 12\penalty0 (7315):\penalty0 1--7,
  2021.
\newblock \doi{10.1038/s41467-021-27554-4}.

\bibitem[Elliott and Brooks(1997{\natexlab{a}})]{Elliott1997}
H~Elliott and Norman~H Brooks.
\newblock {Transfer of nonsorbing solutes to a streambed with bed forms :
  Theory}.
\newblock \emph{Water Resources Research}, 33\penalty0 (1):\penalty0 123--136,
  1997{\natexlab{a}}.
\newblock \doi{https://doi.org/10.1029/96WR02784}.

\bibitem[Elliott and Brooks(1997{\natexlab{b}})]{Elliott1997a}
H~Elliott and Norman~H Brooks.
\newblock {Transfer of nonsorbing solutes to a streambed with bed forms:
  Laboratory experiments}.
\newblock \emph{Water Resources Research}, 33\penalty0 (1):\penalty0 137--151,
  1997{\natexlab{b}}.

\bibitem[Fehlman(1985)]{Fehlman1985}
Henry~Michael Fehlman.
\newblock \emph{{Resistance Components and Velocity Distributions of Open
  Channel Flows Over Bedforms}}.
\newblock PhD thesis, Colorado State University, 1985.

\bibitem[Fox et~al.(2014)Fox, Boano, and Arnon]{Fox2014}
Aryeh Fox, Fulvio Boano, and Shai Arnon.
\newblock {Impact of losing and gaining streamflow conditions on hyporheic
  exchange fluxes induced by dune-shaped bed forms}.
\newblock \emph{Water Resources Research}, 50\penalty0 (3), 2014.
\newblock ISSN 00431397.
\newblock \doi{10.1002/2013WR014333}.

\bibitem[Gillies and Others(2007)]{Gillies2007}
Sean Gillies and Others.
\newblock {Shapely: manipulation and analysis of geometric objects}, 2007.
\newblock URL \url{https://github.com/shapely/shapely}.

\bibitem[Harvey et~al.(1996)Harvey, Wagner, and Bencala]{Harvey1996}
Judson~W Harvey, Brian~J Wagner, and Kenneth~E Bencala.
\newblock {Evaluating the reliability of the stream tracer approach to
  characterize stream-subsurface water exchange}.
\newblock \emph{Water Resources Research}, 32\penalty0 (8):\penalty0
  2441--2451, 1996.

\bibitem[Huang and Chui(2022)]{Huang2022}
Peng Huang and Ting Fong~May Chui.
\newblock {Hyporheic Exchange in a Straight Stream With Alternate Bars}.
\newblock \emph{Water Resources Research}, 58\penalty0 (10):\penalty0 1--25,
  2022.
\newblock ISSN 19447973.
\newblock \doi{10.1029/2022WR032221}.

\bibitem[Huang and Yang(2023)]{Huang2023}
S.~H. Huang and J.~Q. Yang.
\newblock {Impacts of Channel-Spanning Log Jams on Hyporheic Flow}.
\newblock \emph{Water Resources Research}, 59\penalty0 (11):\penalty0 1--15,
  2023.
\newblock ISSN 19447973.
\newblock \doi{10.1029/2023WR035217}.

\bibitem[Janssen et~al.(2012)Janssen, Cardenas, Sawyer, Dammrich, Krietsch, and
  {De Beer}]{Janssen2012}
Felix Janssen, M.~Bayani Cardenas, Audrey~H. Sawyer, Thea Dammrich, Jana
  Krietsch, and Dirk {De Beer}.
\newblock {A comparative experimental and multiphysics computational fluid
  dynamics study of coupled surface-subsurface flow in bed forms}.
\newblock \emph{Water Resources Research}, 48\penalty0 (8):\penalty0 1--16,
  2012.
\newblock ISSN 00431397.
\newblock \doi{10.1029/2012WR011982}.

\bibitem[Jin et~al.(2022)Jin, Yuan, Zhang, Zhang, Chen, Tang, and Li]{Jin2022}
Guangqiu Jin, Haiyu Yuan, Guangming Zhang, Zhongtian Zhang, Chen Chen, Hongwu
  Tang, and Ling Li.
\newblock {Effects of bed geometric characteristics on hyporheic exchange}.
\newblock \emph{Journal of Hydro-environment Research}, 43:\penalty0 1--9,
  2022.
\newblock ISSN 1570-6443.
\newblock \doi{https://doi.org/10.1016/j.jher.2022.05.001}.
\newblock URL
  \url{https://www.sciencedirect.com/science/article/pii/S1570644322000302}.

\bibitem[Kennedy(1969)]{Kennedy1969}
J~F Kennedy.
\newblock {The Formation of Sediment Ripples, Dunes, and Antidunes}.
\newblock \emph{Annual Review of Fluid Mechanics}, 1\penalty0 (1):\penalty0
  147--168, 1969.
\newblock ISSN 0066-4189.
\newblock \doi{10.1146/annurev.fl.01.010169.001051}.

\bibitem[Lee et~al.(2020)Lee, Aubeneau, and Cardenas]{Lee2020}
Anzy Lee, Antoine~F. Aubeneau, and M.~Bayani Cardenas.
\newblock {The Sensitivity of Hyporheic Exchange to Fractal Properties of
  Riverbeds}.
\newblock \emph{Water Resources Research}, 56\penalty0 (5):\penalty0 1--15,
  2020.
\newblock ISSN 19447973.
\newblock \doi{10.1029/2019WR026560}.

\bibitem[Lee et~al.(2022)Lee, Aubeneau, Cardenas, and Liu]{Lee2022}
Anzy Lee, Antoine~F. Aubeneau, M.~Bayani Cardenas, and Xiaofeng Liu.
\newblock {Hyporheic Exchange Due to Cobbles on Sandy Beds}.
\newblock \emph{Water Resources Research}, 58\penalty0 (1):\penalty0 1--14,
  2022.
\newblock ISSN 19447973.
\newblock \doi{10.1029/2021WR030164}.

\bibitem[Li et~al.(2020)Li, Liu, Kaufman, Turetcaia, Chen, and
  Cardenas]{Li2020a}
Bing Li, Xiaofeng Liu, Matthew~H. Kaufman, Anna Turetcaia, Xingyuan Chen, and
  M.~Bayani Cardenas.
\newblock {Flexible and Modular Simultaneous Modeling of Flow and Reactive
  Transport in Rivers and Hyporheic Zones}.
\newblock \emph{Water Resources Research}, 56\penalty0 (2):\penalty0 1--11,
  2020.
\newblock ISSN 19447973.
\newblock \doi{10.1029/2019WR026528}.

\bibitem[Lv et~al.(2022)Lv, Gao, and Sun]{Lv2022}
Jianzhang Lv, Xueping Gao, and Bowen Sun.
\newblock {Numerical simulation of hyporheic exchange driven by an emergent
  vegetation patch}.
\newblock \emph{Hydrological Processes}, 36\penalty0 (11), 2022.
\newblock ISSN 10991085.
\newblock \doi{10.1002/hyp.14756}.

\bibitem[Marion et~al.(2002)Marion, Bellinello, Guymer, and
  Packman]{Marion2002}
Andrea Marion, Matteo Bellinello, Ian Guymer, and Aaron Packman.
\newblock {Effect of bed form geometry on the penetration of nonreactive
  solutes into a streambed}.
\newblock \emph{Water Resources Research}, 38\penalty0 (10):\penalty0
  27--1--27--12, 2002.
\newblock ISSN 00431397.
\newblock \doi{10.1029/2001wr000264}.

\bibitem[Marzadri et~al.(2010)Marzadri, Tonina, Bellin, Vignoli, and
  Tubino]{Marzadri2010}
A.~Marzadri, D.~Tonina, A.~Bellin, G.~Vignoli, and M.~Tubino.
\newblock {Semianalytical analysis of hyporheic flow induced by alternate
  bars}.
\newblock \emph{Water Resources Research}, 46\penalty0 (7):\penalty0 1--14,
  2010.
\newblock ISSN 00431397.
\newblock \doi{10.1029/2009WR008285}.

\bibitem[Raudkivi(1963)]{Raudkivi1963}
A~J Raudkivi.
\newblock {Study of Sediment Ripple Formation}.
\newblock \emph{Journal of the Hydraulics Division}, 89\penalty0 (6):\penalty0
  15--34, 1963.
\newblock \doi{10.1061/JYCEAJ.0000952}.
\newblock URL \url{https://ascelibrary.org/doi/abs/10.1061/JYCEAJ.0000952}.

\bibitem[Risse-Buhl et~al.(2023)Risse-Buhl, Arnon, Bar-Zeev, Oprei, Packman,
  Peralta-Maraver, Robertson, Teitelbaum, and Mutz]{Risse-Buhl2023}
Ute Risse-Buhl, Shai Arnon, Edo Bar-Zeev, Anna Oprei, Aaron~I. Packman, Ignacio
  Peralta-Maraver, Anne Robertson, Yoni Teitelbaum, and Michael Mutz.
\newblock {Streambed migration frequency drives ecology and biogeochemistry
  across spatial scales}.
\newblock \emph{Wiley Interdisciplinary Reviews: Water}, 10\penalty0
  (3):\penalty0 1--13, 2023.
\newblock ISSN 20491948.
\newblock \doi{10.1002/wat2.1632}.

\bibitem[Savant et~al.(1987)Savant, Reible, and Thibodeaux]{Savant1987}
Anne~S. Savant, Danny~D. Reible, and Louis~J. Thibodeaux.
\newblock {Convective transport within stable river sediments}.
\newblock \emph{Water Resources Research}, 23\penalty0 (9):\penalty0
  1763--1768, 1987.

\bibitem[Sawyer et~al.(2011)Sawyer, {Bayani Cardenas}, and Buttles]{Sawyer2011}
Audrey~H. Sawyer, M.~{Bayani Cardenas}, and Jim Buttles.
\newblock {Hyporheic exchange due to channel-spanning logs}.
\newblock \emph{Water Resources Research}, 47\penalty0 (8):\penalty0 1--12,
  2011.
\newblock ISSN 00431397.
\newblock \doi{10.1029/2011WR010484}.

\bibitem[Stonedahl et~al.(2010)Stonedahl, Harvey, W{\"{o}}rman, Salehin, and
  Packman]{Stonedahl2010}
Susa~H. Stonedahl, Judson~W. Harvey, Anders W{\"{o}}rman, Mashfiqus Salehin,
  and Aaron~I. Packman.
\newblock {A multiscale model for integrating hyporheic exchange from ripples
  to meanders}.
\newblock \emph{Water Resources Research}, 46\penalty0 (12):\penalty0 W12539,
  2010.
\newblock ISSN 00431397.
\newblock \doi{10.1029/2009WR008865}.

\bibitem[Teitelbaum et~al.(2021)Teitelbaum, Dallmann, Phillips, Packman,
  Schumer, Sund, Hansen, and Arnon]{Teitelbaum2021}
Yoni Teitelbaum, Jonathan Dallmann, Colin~B. Phillips, Aaron~I. Packman, Rina
  Schumer, Nicole~L. Sund, Scott~K. Hansen, and Shai Arnon.
\newblock {Dynamics of hyporheic exchange flux and fine particle deposition
  under moving bedforms}.
\newblock \emph{Water Resources Research}, 57\penalty0 (4):\penalty0 1--13,
  2021.
\newblock ISSN 0043-1397.
\newblock \doi{10.1029/2020wr028541}.

\bibitem[Teitelbaum et~al.(2022)Teitelbaum, Shimony, {Saavedra Cifuentes},
  Dallmann, Phillips, Packman, Hansen, and Arnon]{Teitelbaum2022}
Yoni Teitelbaum, Tomer Shimony, Edwin {Saavedra Cifuentes}, Jonathan Dallmann,
  Colin~B. Phillips, Aaron~I. Packman, Scott~K. Hansen, and Shai Arnon.
\newblock {A novel framework for simulating particle deposition with moving
  bedforms}.
\newblock \emph{Geophysical Research Letters}, 49\penalty0 (4), 2022.
\newblock ISSN 19448007.
\newblock \doi{10.1029/2021GL097223}.

\bibitem[Thibodeaux and Boyle(1987)]{Thibodeaux1987a}
Louis~J. Thibodeaux and John~D. Boyle.
\newblock {Bedform-generated convective transport in bottom sediment}.
\newblock \emph{Nature}, 325\penalty0 (6102):\penalty0 341--343, jan 1987.
\newblock URL \url{http://dx.doi.org/10.1038/325341a0}.

\bibitem[Toth(1963)]{Toth1963}
J~Toth.
\newblock {A theoretical analysis of ground water flow in small drainage
  basins}.
\newblock \emph{Journal of Geophysical Research}, 68\penalty0 (16):\penalty0
  4795--4812, 1963.

\bibitem[van~der Walt et~al.(2014)van~der Walt, Sch{\"{o}}nberger,
  Nunez-Iglesias, Boulogne, Warner, Yager, Gouillart, Yu, and scikit-image
  contributors]{scikit-image}
St{\'{e}}fan van~der Walt, Johannes~L Sch{\"{o}}nberger, Juan Nunez-Iglesias,
  Fran{\c{c}}ois Boulogne, Joshua~D Warner, Neil Yager, Emmanuelle Gouillart,
  Tony Yu, and The scikit-image contributors.
\newblock {scikit-image: image processing in Python}.
\newblock \emph{PeerJ}, 2:\penalty0 e453, 2014.
\newblock ISSN 2167-8359.
\newblock \doi{10.7717/peerj.453}.
\newblock URL \url{https://doi.org/10.7717/peerj.453}.

\bibitem[van Mierlo and de~Ruiter(1988)]{vanMierlo1988}
M.C.L.M. van Mierlo and J.C.C. de~Ruiter.
\newblock {Turbulence Measurements Above Artificial Dunes}.
\newblock Technical Report Q789, Delft Hydraulics, Delft, Netherlands, 1988.

\bibitem[Vanoni and Hwang(1967)]{Vanoni1967}
Vito~A Vanoni and L~S Hwang.
\newblock {Bed Forms and Friction in Streams}.
\newblock \emph{Journal of Hydraulic Engineering}, 1967.
\newblock URL \url{https://api.semanticscholar.org/CorpusID:127244809}.

\bibitem[Vaux(1968)]{Vaux1968}
W~G Vaux.
\newblock {Intergravel Flow and Exchange of Water in a Streambed}.
\newblock \emph{Fishery Bulletin}, 66\penalty0 (3):\penalty0 479--489, 1968.

\bibitem[Virtanen et~al.(2020)Virtanen, Gommers, Oliphant, Haberland, Reddy,
  Cournapeau, Burovski, Peterson, Weckesser, Bright, van~der Walt, Brett,
  Wilson, Millman, Mayorov, Nelson, Jones, Kern, Larson, Carey, Polat, Feng,
  Moore, VanderPlas, Laxalde, Perktold, Cimrman, Henriksen, Quintero, Harris,
  Archibald, Ribeiro, Pedregosa, van Mulbregt, Vijaykumar, Bardelli, Rothberg,
  Hilboll, Kloeckner, Scopatz, Lee, Rokem, Woods, Fulton, Masson,
  H{\"{a}}ggstr{\"{o}}m, Fitzgerald, Nicholson, Hagen, Pasechnik, Olivetti,
  Martin, Wieser, Silva, Lenders, Wilhelm, Young, Price, Ingold, Allen, Lee,
  Audren, Probst, Dietrich, Silterra, Webber, Slavi{\v{c}}, Nothman, Buchner,
  Kulick, Sch{\"{o}}nberger, {de Miranda Cardoso}, Reimer, Harrington,
  Rodr{\'{i}}guez, Nunez-Iglesias, Kuczynski, Tritz, Thoma, Newville,
  K{\"{u}}mmerer, Bolingbroke, Tartre, Pak, Smith, Nowaczyk, Shebanov, Pavlyk,
  Brodtkorb, Lee, McGibbon, Feldbauer, Lewis, Tygier, Sievert, Vigna, Peterson,
  More, Pudlik, Oshima, Pingel, Robitaille, Spura, Jones, Cera, Leslie, Zito,
  Krauss, Upadhyay, Halchenko, and V{\'{a}}zquez-Baeza]{2020SciPy-NMeth}
Pauli Virtanen, Ralf Gommers, Travis~E. Oliphant, Matt Haberland, Tyler Reddy,
  David Cournapeau, Evgeni Burovski, Pearu Peterson, Warren Weckesser, Jonathan
  Bright, St{\'{e}}fan~J. van~der Walt, Matthew Brett, Joshua Wilson, K.~Jarrod
  Millman, Nikolay Mayorov, Andrew~R.J. Nelson, Eric Jones, Robert Kern, Eric
  Larson, C.~J. Carey, İlhan Polat, Yu~Feng, Eric~W. Moore, Jake VanderPlas,
  Denis Laxalde, Josef Perktold, Robert Cimrman, Ian Henriksen, E.~A. Quintero,
  Charles~R. Harris, Anne~M. Archibald, Ant{\^{o}}nio~H. Ribeiro, Fabian
  Pedregosa, Paul van Mulbregt, Aditya Vijaykumar, Alessandro~Pietro Bardelli,
  Alex Rothberg, Andreas Hilboll, Andreas Kloeckner, Anthony Scopatz, Antony
  Lee, Ariel Rokem, C.~Nathan Woods, Chad Fulton, Charles Masson, Christian
  H{\"{a}}ggstr{\"{o}}m, Clark Fitzgerald, David~A. Nicholson, David~R. Hagen,
  Dmitrii~V. Pasechnik, Emanuele Olivetti, Eric Martin, Eric Wieser, Fabrice
  Silva, Felix Lenders, Florian Wilhelm, G.~Young, Gavin~A. Price, Gert~Ludwig
  Ingold, Gregory~E. Allen, Gregory~R. Lee, Herv{\'{e}} Audren, Irvin Probst,
  J{\"{o}}rg~P. Dietrich, Jacob Silterra, James~T. Webber, Janko Slavi{\v{c}},
  Joel Nothman, Johannes Buchner, Johannes Kulick, Johannes~L.
  Sch{\"{o}}nberger, Jos{\'{e}}~Vin{\'{i}}cius {de Miranda Cardoso}, Joscha
  Reimer, Joseph Harrington, Juan Luis~Cano Rodr{\'{i}}guez, Juan
  Nunez-Iglesias, Justin Kuczynski, Kevin Tritz, Martin Thoma, Matthew
  Newville, Matthias K{\"{u}}mmerer, Maximilian Bolingbroke, Michael Tartre,
  Mikhail Pak, Nathaniel~J. Smith, Nikolai Nowaczyk, Nikolay Shebanov,
  Oleksandr Pavlyk, Per~A. Brodtkorb, Perry Lee, Robert~T. McGibbon, Roman
  Feldbauer, Sam Lewis, Sam Tygier, Scott Sievert, Sebastiano Vigna, Stefan
  Peterson, Surhud More, Tadeusz Pudlik, Takuya Oshima, Thomas~J. Pingel,
  Thomas~P. Robitaille, Thomas Spura, Thouis~R. Jones, Tim Cera, Tim Leslie,
  Tiziano Zito, Tom Krauss, Utkarsh Upadhyay, Yaroslav~O. Halchenko, and
  Yoshiki V{\'{a}}zquez-Baeza.
\newblock {SciPy 1.0: fundamental algorithms for scientific computing in
  Python}.
\newblock \emph{Nature Methods}, 17\penalty0 (3):\penalty0 261--272, 2020.
\newblock ISSN 15487105.
\newblock \doi{10.1038/s41592-019-0686-2}.

\bibitem[Vittal et~al.(1977)Vittal, Raju, and Garde]{Vittal1977}
N.~Vittal, K.~G.Ranga Raju, and R.~J. Garde.
\newblock {Resistance of Two Dimensional Triangular Roughness}.
\newblock \emph{Journal of Hydraulic Research}, 15\penalty0 (1):\penalty0
  19--36, 1977.
\newblock ISSN 00221686.
\newblock \doi{10.1080/00221687709499747}.

\bibitem[Winter et~al.(1998)Winter, {Judson W. Harvey}, Franke, and
  Alley]{Winter1998}
Thomas~C. Winter, {Judson W. Harvey}, O.~Lehn Franke, and William~M. Alley.
\newblock {Ground Water and Surface Water: A Single Resource}.
\newblock \emph{United States Geological Survey Circular}, 1139, 1998.

\bibitem[W{\"{o}}rman et~al.(2006)W{\"{o}}rman, Packman, Marklund, Harvey, and
  Stone]{Worman2006}
Anders W{\"{o}}rman, Aaron~I. Packman, Lars Marklund, Judson~W. Harvey, and
  Susa~H. Stone.
\newblock {Exact three-dimensional spectral solution to surface-groundwater
  interactions with arbitrary surface topography}.
\newblock \emph{Geophysical Research Letters}, 33\penalty0 (7):\penalty0 2--5,
  2006.
\newblock ISSN 00948276.
\newblock \doi{10.1029/2006GL025747}.

\bibitem[W{\"{o}}rman et~al.(2007)W{\"{o}}rman, Packman, Marklund, Harvey, and
  Stone]{Worman2007}
Anders W{\"{o}}rman, Aaron~I. Packman, Lars Marklund, Judson~W. Harvey, and
  Susa~H. Stone.
\newblock {Fractal topography and subsurface water flows from fluvial bedforms
  to the continental shield}.
\newblock \emph{Geophysical Research Letters}, 34\penalty0 (7), apr 2007.
\newblock ISSN 0094-8276.
\newblock \doi{10.1029/2007GL029426}.

\bibitem[Yuan et~al.(2021)Yuan, Chen, Cardenas, Liu, and Chen]{Yuan2021}
Yue Yuan, Xiaobing Chen, M.~Bayani Cardenas, Xiaofeng Liu, and Li~Chen.
\newblock {Hyporheic Exchange Driven by Submerged Rigid Vegetation: A Modeling
  Study}.
\newblock \emph{Water Resources Research}, 57\penalty0 (6):\penalty0 1--15,
  2021.
\newblock ISSN 19447973.
\newblock \doi{10.1029/2019WR026675}.

\end{thebibliography}

\end{document}